\documentclass[twocolumn]{aastex63}

\usepackage{natbib}
\usepackage{booktabs}
\usepackage{appendix}

\newcommand{\tess}{\emph{TESS}}

\submitjournal{\aj}
\graphicspath{{./}{figures/}}

\shorttitle{Young quadruple system identified by \tess{}}
\shortauthors{Rowden et al.}

\begin{document}

\title{TIC 278956474: Two close binaries in one young quadruple system, identified by \textit{TESS}}

\correspondingauthor{Pamela~Rowden}
\email{pam.rowden27@gmail.com}

\author[0000-0002-4829-7101]{Pamela~Rowden}
\affiliation{School of Physical Sciences, The Open University, Milton Keynes MK7 6AA, UK}

\author[0000-0002-8806-496X]{Tam\'as~Borkovits}
\affiliation{Baja Astronomical Observatory of Szeged University, H-6500 Baja, Szegedi út, kt. 766, Hungary}
\affiliation{Konkoly Observatory, Research Centre for Astronomy and Earth Sciences, H-1121 Budapest, Konkoly Thege Miklós út 15-17, Hungary}

\author[0000-0002-4715-9460]{Jon~M.~Jenkins}
\affiliation{NASA Ames Research Center, Moffett Field, CA, 94035, USA}

\author[0000-0002-3481-9052]{Keivan~G.~Stassun} 
\affiliation{Department of Physics and Astronomy, Vanderbilt University, Nashville, TN 37235, USA}
\affiliation{Department of Physics, Fisk University, Nashville, TN 37208, USA}

\author[0000-0002-6778-7552]{Joseph~D.~Twicken}
\affiliation{NASA Ames Research Center, Moffett Field, CA, 94035, USA}
\affiliation{SETI Institute, Mountain View, CA 94043, USA}

\author[0000-0003-4150-841X]{Elisabeth~R.~Newton}
\affiliation{Department of Physics and Astronomy, Dartmouth College, Hanover, NH, 03755, USA}

\author[0000-0002-0619-7639]{Carl Ziegler}
\affil{Dunlap Institute for Astronomy and Astrophysics, University of Toronto, 50 St. George Street, Toronto, Ontario M5S 3H4, Canada}

\author[0000-0002-3439-1439]{Coel~Hellier}
\affiliation{Astrophysics Group, Keele University, Staffordshire, ST5 5BG, UK}

\author[0000-0001-9828-3229]{Aylin~Garcia~Soto}
\affiliation{Department of Physics and Astronomy, Dartmouth College, Hanover, NH, 03755, USA}

\author[0000-0003-0593-1560]{Elisabeth~C.~Matthews}
\affiliation{Department of Physics and Kavli Institute for Astrophysics and Space Research, Massachusetts Institute of Technology, Cambridge, MA 02139, USA}

\author[0000-0001-8670-8365]{Ulrich~Kolb}
\affiliation{School of Physical Sciences, The Open University, Milton Keynes MK7 6AA, UK}

\author[0000-0003-2058-6662]{George~R.~Ricker}
\affiliation{Department of Physics and Kavli Institute for Astrophysics and Space Research, Massachusetts Institute of Technology, Cambridge, MA 02139, USA}

\author[0000-0001-6763-6562]{Roland~Vanderspek}
\affiliation{Department of Physics and Kavli Institute for Astrophysics and Space Research, Massachusetts Institute of Technology, Cambridge, MA 02139, USA}

\author[0000-0001-9911-7388]{David~W.~Latham}
\affil{Center for Astrophysics ${\rm \mid}$ Harvard {\rm \&} Smithsonian, 60 Garden Street, Cambridge, MA 02138, USA}

\author[0000-0002-6892-6948]{S.~Seager}
\affiliation{Department of Physics and Kavli Institute for Astrophysics and Space Research, Massachusetts Institute of Technology, Cambridge, MA 02139, USA}

\affiliation{Department of Earth, Atmospheric and Planetary Sciences, Massachusetts Institute of Technology, Cambridge, MA 02139, USA}
\affiliation{Department of Aeronautics and Astronautics, MIT, 77 Massachusetts Avenue, Cambridge, MA 02139, USA}

\author[0000-0002-4265-047X]{Joshua~N.~Winn}
\affiliation{Department of Astrophysical Sciences, Princeton University, 4 Ivy Lane, Princeton, NJ 08544, USA}

\author[0000-0002-0514-5538]{Luke~G.~Bouma}
\affiliation{Department of Astrophysical Sciences, Princeton University, 4 Ivy Lane, Princeton, NJ 08544, USA}

\author[0000-0001-7124-4094]{C\'{e}sar Brice\~{n}o}
\affiliation{Cerro Tololo Inter-American Observatory, Casilla 603, La Serena, Chile} 

\author[0000-0002-9003-484X]{David~Charbonneau}
\affil{Center for Astrophysics ${\rm \mid}$ Harvard {\rm \&} Smithsonian, 60 Garden Street, Cambridge, MA 02138, USA}

\author[0000-0003-0241-2757]{William~Fong}
\affiliation{Department of Physics and Kavli Institute for Astrophysics and Space Research, Massachusetts Institute of Technology, Cambridge, MA 02139, USA}

\author[0000-0002-5322-2315]{Ana~Glidden}
\affiliation{Department of Earth, Atmospheric and Planetary Sciences, Massachusetts Institute of Technology, Cambridge, MA 02139, USA}
\affiliation{Department of Physics and Kavli Institute for Astrophysics and Space Research, Massachusetts Institute of Technology, Cambridge, MA 02139, USA}

\author[0000-0002-5169-9427]{Natalia~M.~Guerrero}
\affiliation{Department of Physics and Kavli Institute for Astrophysics and Space Research, Massachusetts Institute of Technology, Cambridge, MA 02139, USA}

\author{Nicholas Law}
\affiliation{Department of Physics and Astronomy, The University of North Carolina at Chapel Hill, Chapel Hill, NC 27599-3255, USA}

\author[0000-0003-3654-1602]{Andrew W. Mann}
\affiliation{Department of Physics and Astronomy, The University of North Carolina at Chapel Hill, Chapel Hill, NC 27599-3255, USA}

\author[0000-0003-4724-745X]{Mark~E.~Rose}
\affiliation{NASA Ames Research Center, Moffett Field, CA, 94035, USA}

\author{Joshua~Schlieder}
\affiliation{NASA Goddard Space Flight Center, 8800 Greenbelt Rd, Greenbelt, MD 20771, USA}

\author[0000-0002-1949-4720]{Peter Tenenbaum}
\affiliation{SETI Institute, Mountain View, CA 94043, USA}
\affiliation{NASA Ames Research Center, Moffett Field, CA, 94035, USA}

\author[0000-0002-8219-9505]{Eric B. Ting}
\affiliation{NASA Ames Research Center, Moffett Field, CA, 94035, USA}

\begin{abstract}
	We have identified a quadruple system with two close eclipsing binaries in \textit{TESS} data. The object is unresolved in Gaia and appears as a single source at parallax 1.08~$\pm$0.01 mas. Both binaries have observable primary and secondary eclipses and were monitored throughout \textit{TESS} Cycle 1 (sectors 1-13), falling within the \textit{TESS} Continuous Viewing Zone. In one eclipsing binary (\textit{P} = 5.488 d), the smaller star is completely occluded by the larger star during the secondary eclipse; in the other (\textit{P} = 5.674 d) both eclipses are grazing. Using these data, spectroscopy, speckle photometry, SED analysis and evolutionary stellar tracks, we have constrained the masses and radii of the four stars in the two eclipsing binaries. The Li I EW indicates an age of 10-50 Myr and, with an outer period of $858^{+7}_{-5}$ days, our analysis indicates this is one of the most compact young 2+2 quadruple systems known.
\end{abstract}


\section{Introduction}
\label{Intro}

The main purpose of the  Transiting Exoplanet Survey Satellite, \textit{TESS} \citep{2014AAS...22411302R}, is to identify nearby planets $\le$ 4 R$_\oplus$ which can be fully characterised. However, a great deal of complementary science has come from the mission, particularly in stellar science (see for example \citet{2019arXiv190402171F}, \citet{2019ApJ...883..111H}, \citet{2019AJ....157..245H, 2019ApJS..241...12S}, \citet{2019ApJ...876..127Z}, \citet{2020ApJ...888...63A}).

Eclipsing binaries are known to be detected in transiting exoplanet surveys. $\approx 16\%$ of \textit{Kepler} Objects of Interest (KOIs) have been identified by the \textit{Kepler} pipeline as eclipsing binaries, and a further $\approx$7\% as background eclipsing binaries\footnote{\url{https://exoplanetarchive.ipac.caltech.edu/docs/data.html}}. Moreover, the \textit{Kepler} and \textit{K2} missions have also identified triple and quadruple eclipsing systems and even a bound quintuple system (KOI 3156) exhibiting eclipses of at least three different subsystems \citep{2017A&A...602A..30H}.

\citet{2010ApJS..190....1R} estimate that, among solar-type stars, 33$\pm2\%$ of systems are binary, 8$\pm1\%$ of systems are triple, 9$\pm2\%$ of systems are quadruple and 3$\pm1\%$ are composed of five stars. As an all-sky survey, \textit{TESS} can be expected to identify a proportion of these rarer multiple star systems.

We use \textit{TESS} data to identify a 2+2 quadruple star system (TIC 278956474) with two short-period inner binaries. We estimate the age of TIC 278956474 as 10-50 Myr (Section~\ref{Spectroscopy}), making this a young system. Known young quadruple systems include GG Tauri \citep{1999A&A...348..570G, 2011A&A...530A.126K}, HD 98800 \citep{1999AstL...25..669T, 2018ApJ...865...77R}, HD 34700 \citep{2005A&A...434..671S}, AB Doradus \citep{2007A&A...462..615J, 2014A&A...570A..95W}, AO Vel \citep{2006A&A...449..327G, 2008psa..conf..291G, 2008msah.conf..259G}, HD 91962 \citep{2015AJ....149..195T}, LkCa 3 \citep{2013ApJ...773...40T, 2015A&A...577A..42B} and HD 86588 \citep{2018AJ....156..120T}. IRS5 might be a young quadruple system \citep{2015A&A...578A..82C}. LkH$\alpha$ 263C, around which a circumstellar disc has been identified \citep{2002ApJ...571L..51J}, appears to be a member of a young quadruple system in the MBM 12 association. 

While considering the binary population of young clusters, \citet{2012A&A...543A...8M} demonstrated from simulations that clusters with a formal binary fraction of unity at birth will evolve to a lower binary fraction over time. They note that younger clusters appear to have a higher binary fraction than older clusters with a similar stellar density (see for example \citet{1999A&A...341..547D}). Thus, studying the population of young quadruple systems such as TIC 278956474 is of interest when considering the evolution of the binary fraction of stellar clusters over time.

This paper is organised as follows. In Section~\ref{Methods} we discuss the data. In Section~\ref{Models} we present our models, which confirm that this is a young 2+2 quadruple system. The models are discussed in Section~\ref{Discussion}: in particular, we consider the dynamical properties of the system, as well as its place among known young quadruple star systems. Our conclusions are drawn in Section~\ref{Conclusion}.

\section{Data}
\label{Methods}

\begin{table}
\caption{Basic data on TIC 278956474. }
\centering
\label{table:parameters}
\scalebox{1.}{
\begin{tabular}{l l}
\hline
Parameter & Value\\
\hline
Alternative names$^1$ & UCAC4 165-008872\\
& 2MASS J06454123-5708171\\
& WISE J064541.25-570817.0\\
& APASS 27316174\\
\hline
RA$^{1,2}$ & 101.421895$^\circ$\\
dec$^{1,2}$ & -57.138098$^\circ$\\
\textit{l}$^1$ & 266.7396$^\circ$\\
\textit{b}$^1$ & -23.2743$^\circ$\\
\hline
Parallax$^2$ & 1,08 $\pm$ 0.01 mas\\
Proper motion RA$^2$ & 4.29 $\pm$ 0.03 mas yr$^{-1}$\\
Proper motion dec$^2$ & -2.21 $\pm$ 0.03 mas yr$^{-1}$\\
\hline
B$^1$ & 14.191 $\pm$ 0.052\\
\textit{Gaia} bp$^2$ & 13.7641\\ 
V$^1$ & 13.542 $\pm$ 0.092\\
\textit{Gaia}$^{1,2}$ & 13.4153 $\pm$ 0.000408\\
\textit{TESS}$^1$ & 12.9637 $\pm$ 0.006\\
\textit{Gaia} rp$^2$ & 12.9020\\
J$^1$ & 12.291 $\pm$ 0.022\\
H$^1$ & 11.951 $\pm$ 0.024\\
K$^1$ & 11.835 $\pm$ 0.021\\
WISE 3.4 micron$^1$ & 11.813 $\pm$ 0.023\\
WISE 4.6 micron$^1$ & 11.826 $\pm$ 0.021\\
WISE 12 micron$^1$ & 12.048 $\pm$ 0.178\\
WISE 22 micron$^1$ & 9.681\\
\hline
\\
\end{tabular}}

Sources: 1.~Exofop \url{https://exofop.ipac.caltech.edu/tess/}. 2.~\textit{Gaia} DR2 \url{https://gea.esac.esa.int/archive/}.
\end{table}

Table~\ref{table:parameters} gives some basic data on TIC 278956474, such as alternative names, position, proper motion and magnitudes in various passbands. The data are drawn from Exofop\footnote{\url{https://exofop.ipac.caltech.edu/tess/}}, and from \textit{Gaia} DR2\footnote{\url{https://gea.esac.esa.int/archive/}}.

\subsection{SPOC data}
\label{SPOC}

Threshold crossing events (TCEs) were identified in observations of TIC 278956474 in two minute cadence data, processed by NASA's TESS Science Processing Operations Center (SPOC) \citep{2019AAS...23342302J, 2016SPIE.9913E..3EJ}. TIC 278956474 lies in the Southern Continuous Viewing Zone (CVZ) near the Southern ecliptic pole and was observed on camera 4 throughout \textit{TESS} Cycle 1 (Sectors 1-13). We focus on the depth of each SPOC TCE in ppm.

\begin{figure}
\centering
	\includegraphics[width=\columnwidth]{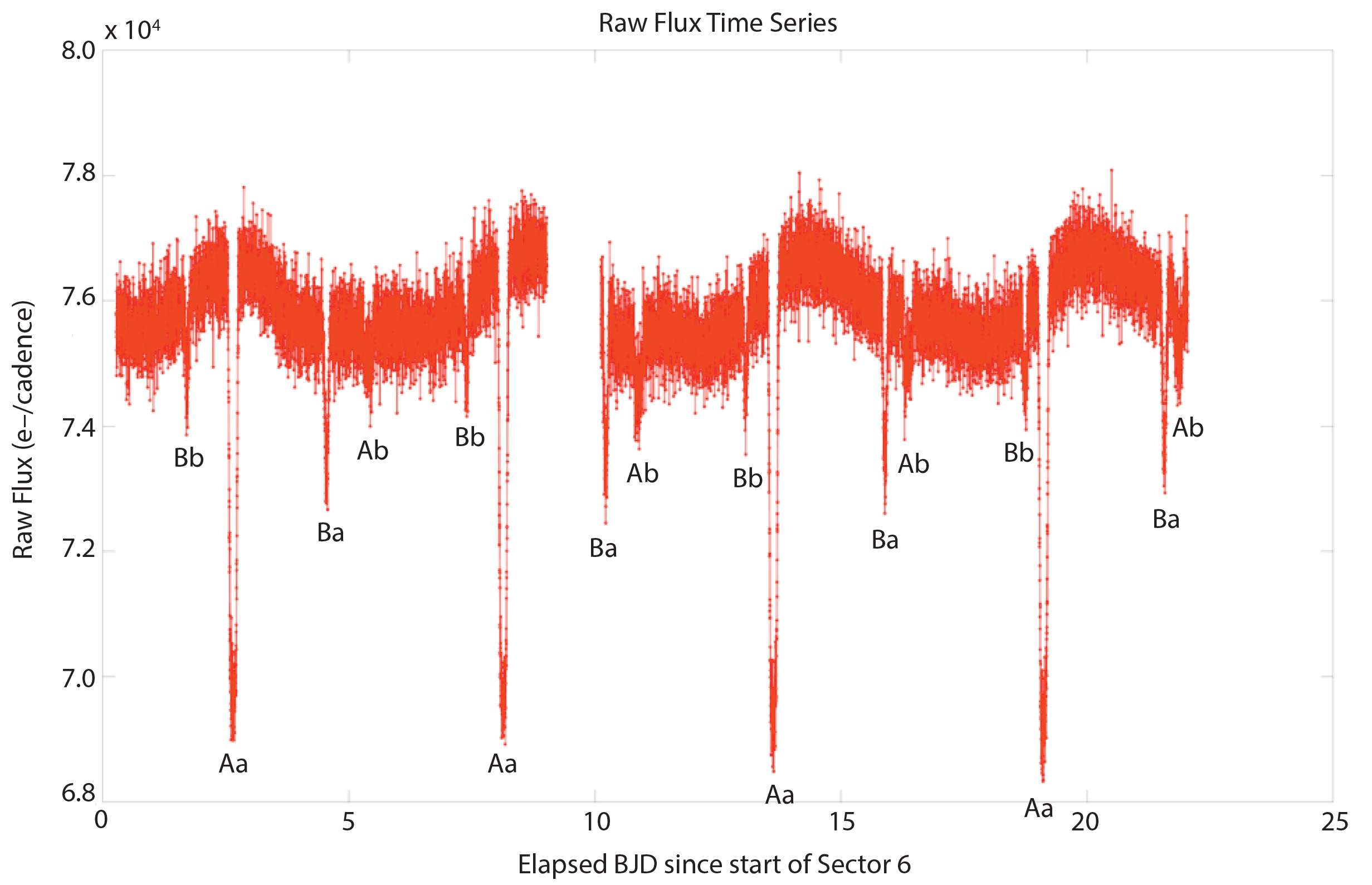}
    \caption{Simple Aperture Photometry (SAP) light curve for Sector 6, annotated to indicate the various eclipses. The start BJD for Sector 6 is 2458468. Similar raw flux light curves are available for each \textit{TESS} sector in the range $1-13$ and are included in the publicly available data validation reports hosted on MAST.}
    \label{fig:Raw}
\end{figure}

Fig.~\ref{fig:Raw} illustrates the Simple Aperture Photometry (SAP) light curve \citep{2010SPIE.7740E..23T, 2017ksci.rept....6M} for sector 6, annotated to highlight the eclipses. Similar information is available on all 13 sectors, and is included in the publicly available data validation reports hosted on MAST.

Selected data on this target from SPOC data validation reports (\citet{2018PASP..130f4502T}, \citet{2019PASP..131b4506L}) is presented in Table~\ref{table:SPOC1-13} (multi-sector analysis, sectors 1-13) and in Table~\ref{table:SPOC} (all single sector and multi sector analyses).

The deepest eclipse and the shallowest eclipse both relate to the same binary (`A'). This binary has a period of 5.488 days. We label the two components Aa and Ab. The second binary (`B') has a period of 5.674 days, and its components are labelled Ba and Bb.

\begin{figure}
\centering
	\includegraphics[width=.7\columnwidth]{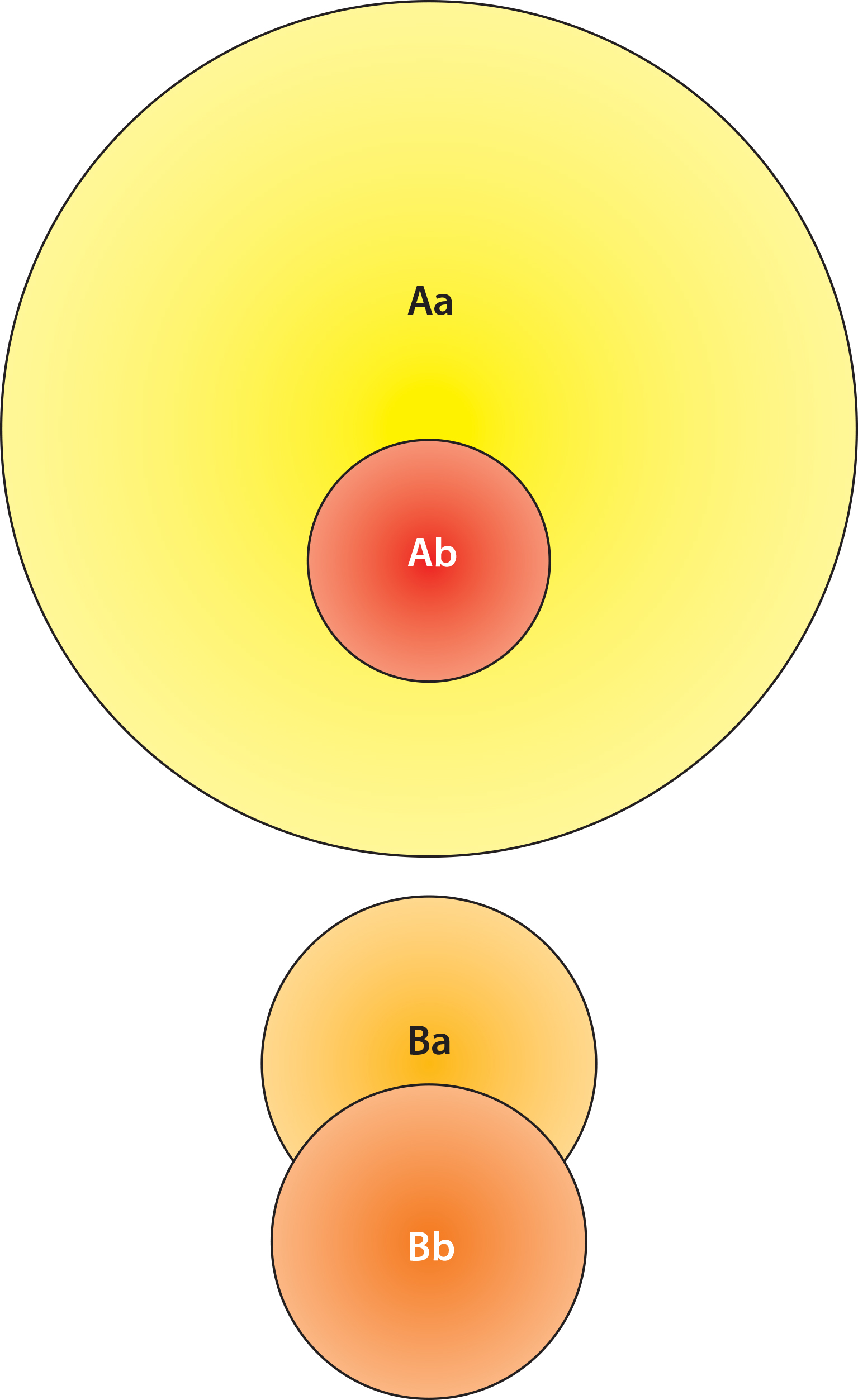}
    \caption{Cartoon showing the relative sizes of the four stars to scale and the proportions of the stars that are occluded during eclipses. Each pair of stars is positioned as it would be at the middle of the primary transit, given the approximate angle of inclination, as observed by TESS.}
    \label{fig:Cartoon}
\end{figure}

The SPOC analysis indicates that both eclipses in B are V-shaped, while both eclipses in A are U-shaped. This indicates the eclipses in B are grazing, while in A star Ab is fully occluded as it passes behind Aa. We obtain a preliminary estimate of the ratio of the radii of Ab:Aa ($\approx$ 0.29) by comparing the ingress duration with the total eclipse duration. See Fig.~\ref{fig:Cartoon} for a cartoon illustrating the relative radii of the four stars to scale and the proportion of each star that is occluded during an eclipse. Each pair of stars is positioned as it would be at the middle of the primary transit, given the approximate angle of inclination, as observed by TESS.

\begin{table*}
\caption{Data from SPOC. The secondary in component Aa coincides with the time of the primary in component Ab, and vice versa. Similarly for components Ba and Bb. Secondary depths are therefore only included where each component is not separately identified, as is the case with B in this table, which has been identified at half the true period. A second science run was completed for multisector 1-13 in order to identify component Ab at the correct period and to remove the partial eclipse of Aa at the end of sector 8. This science run is identified as sector 1-13*. For data from each sector, see Table~\ref{table:SPOC}.}
\centering
\label{table:SPOC1-13}
\scalebox{0.7}{
\begin{tabular}{l l l l l l l l l}
\toprule
Component & Sector & Period/days & Depth/ppm & Duration/hr & Ingress/hr & Odd depth/ppm & Even depth/ppm & Secondary depth/ppm\\
\midrule
Aa & 1-13 & 5.488035 $\pm$ 0.000005 & 93900 $\pm$ 200 & 5.43 $\pm$ 0.01 & 1.22 $\pm$ 0.01 & 94600 $\pm$ 300 & 93300 $\pm$ 300 & 8900 $\pm$ 400\\
Aa & 1-13* & 5.488036 $\pm$ 0.000005 & 93900 $\pm$ 200 & 5.43 $\pm$ 0.01 & 1.22 $\pm$ 0.01 & 94500 $\pm$ 300 & 93400 $\pm$ 300 & Component Ab\\
\midrule
Ab & 1-13 & 2.74403 $\pm$ 0.00002 & 8900 $\pm$ 200 & 5.29 $\pm$ 0.09 & 1.0 $\pm$ 0.1 & Model fitter failed & Model fitter failed & n/a\\
Ab & 1-13* & 5.48808 $\pm$ 0.00004 & 8900 $\pm$ 200 & 5.29 $\pm$ 0.08 & 1.0 $\pm$ 0.1 & 9000 $\pm$ 300 & 8700 $\pm$ 300 & Component Aa\\
\midrule
B & 1-13 & 2.837163 $\pm$ 0.000007 & 25100 $\pm$ 200 & 3.34 $\pm$ 0.03 & 1.67 $\pm$ 0.02 & 16400 $\pm$ 300 & 33400 $\pm$ 300 & n/a\\
B & 1-13* & 2.837162 $\pm$ 0.000007 & 25100 $\pm$ 200 & 3.34 $\pm$ 0.03 & 1.67 $\pm$ 0.02 & 16400 $\pm$ 200 & 33400 $\pm$ 300 & n/a\\
\bottomrule
\end{tabular}}
\end{table*}

\subsection{WASP-South photometry}
\label{WASP}

WASP-South was the southern station of the WASP transit-search project \citep{2006PASP..118.1407P}, situated in Sutherland, South Africa. It observed the field of TIC 278956474 for four consecutive years from 2008 September, spanning 170 nights each year, and obtaining a total of 26\,700 photometric data points. The observations used 200-mm, f/1.8 lenses with a 400--700 nm passband, backed by $2048 \times 2048$ CCDs. Reduction with the WASP pipeline produced photometry relative to other stars in the field with an extraction aperture of 48". At a {\it Gaia\/} magnitude of 13.4, the star is at the faint end of the WASP range, but the data are sufficient to detect 10\%\ eclipses. 

\subsection{Period study}
\label{ETV}

One way to determine whether two eclipsing binaries producing a blended lightcurve are physically bound is to find anti-correlated eclipse timing variations (ETV) in the two pairs. Similar anti-correlated ETVs have proven the real, bound 2+2 quadruple nature of V994~Her \citep{2008MNRAS.389.1630L,2016A&A...588A.121Z} and EPIC~220204960 \citep{2017MNRAS.467.2160R}. More recently, \citet{2019A&A...630A.128Z} performed a thorough analysis of a larger sample of doubly eclipsing binaries found within the frame of the several year-long photometry of the Optical Gravitational Lensing Experiment (OGLE) survey \citep{2015AcA....65....1U}, and identified 28 systems where the ETVs showed evidence of light-travel time effect caused by the relative motion of the two binaries around their common center of mass and/or perturbations due to the dynamical interactions of the two binaries. To search for ETVs in the two eclipsing binaries in TIC 278956474, we determined the times of minimum light of each eclipse  observed with \textit{TESS} in the same manner as was described in Section~5 of \citet{2018MNRAS.478.5135B}.

\begin{figure*}
\begin{center}
\includegraphics[width=0.49 \textwidth]{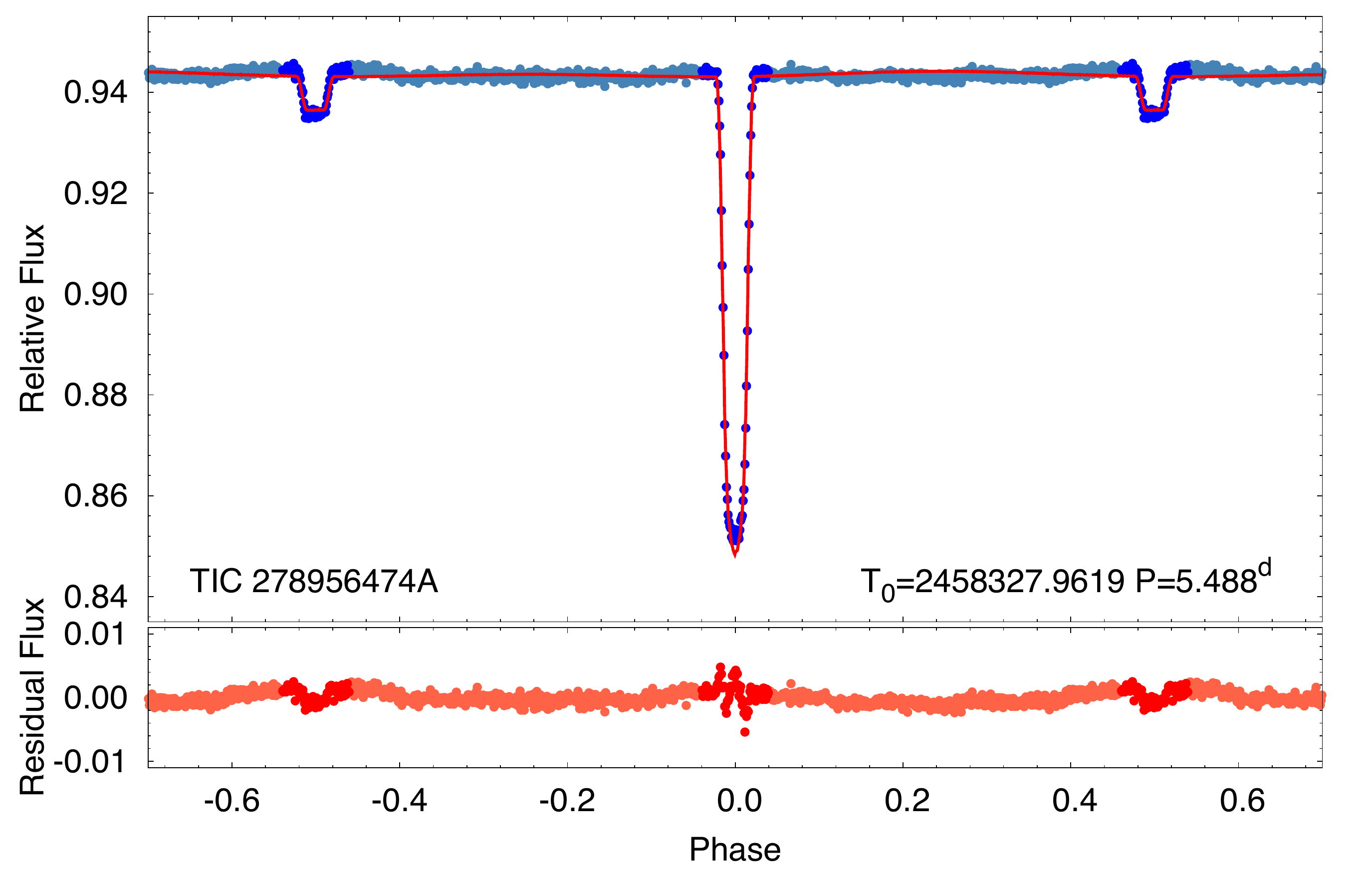}
\includegraphics[width=0.49 \textwidth]{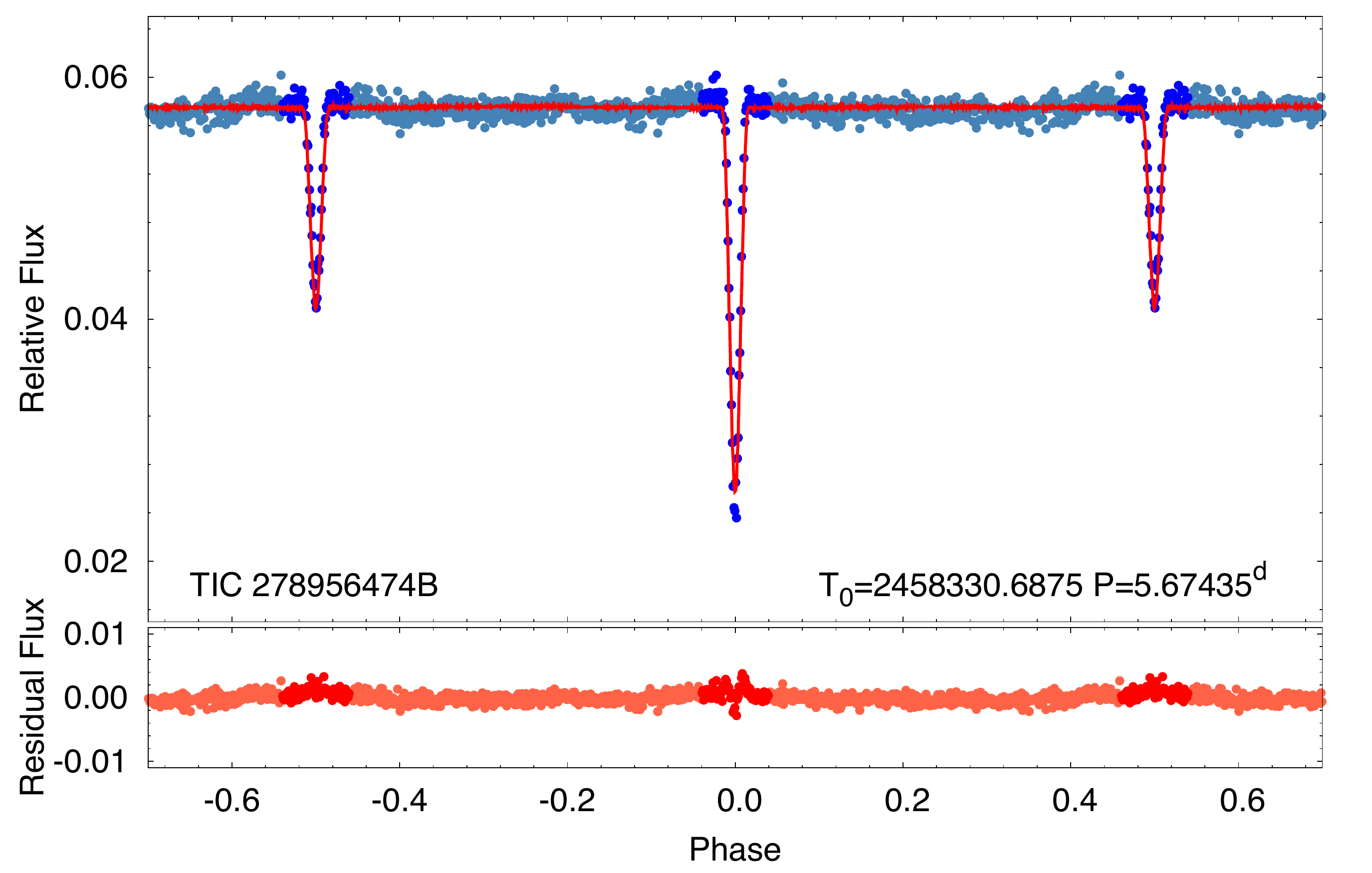}
\caption{The disentangled, phase-folded, binned, averaged light curves (blue points) of binaries A (left) and B (right), together with the complex model solution light curves (see below, in Sect.\,\ref{lcRVETVanalysis}), processed in the same manner. (For the joint analysis only the darker blue points were used.) The lowest, residual data were also obtained with phase-folding, binning and averaging the residual curve of the complete light curve model.}
\label{fig:lcfold} 
\end{center}
\end{figure*}

In summary, after removing the eclipses of the other binary the light curves were phase folded, binned into 1000 equally phased cells, and averaged within each cells. In this way we obtained distentangled, phase folded light curves for both binaries (see Fig.\,\ref{fig:lcfold}). Then, the eclipses of these light curves were fitted with 8--10$^{th}$ order polynoms, and in this way we obtained separate templates for both the primary and secondary eclipses. Then these templates were fitted to each individual eclipse events. (Naturally, we excluded those events which were affected by any eclipses of the other binary.) We obtained $\sim4\times50$ separate minima times (primary and secondary eclipses for both systems) (Tables~\ref{Tab:TIC_278956474A_ToM} and \ref{Tab:TIC_278956474B_ToM}). In what follows, however, we concentrate only on the ETVs of the two primary eclipses, as the secondary ETV points, determined from shallower eclipses, have much higher scatter.

We also took into account the historical WASP-South observations (see Section~\ref{WASP}). These data have large scatter, and therefore, are unsuitable for determining individual eclipse times. However, folding these measurements with the period of binary A season by season, we were able to determine additional seasonal primary minimum times for binary A with a reasonable accuracy. These four seasonal minima are also tabulated in Tables~\ref{Tab:TIC_278956474A_ToM} and \ref{Tab:TIC_278956474B_ToM}.

\begin{figure*}
\centering
        \includegraphics[width=0.49 \textwidth]{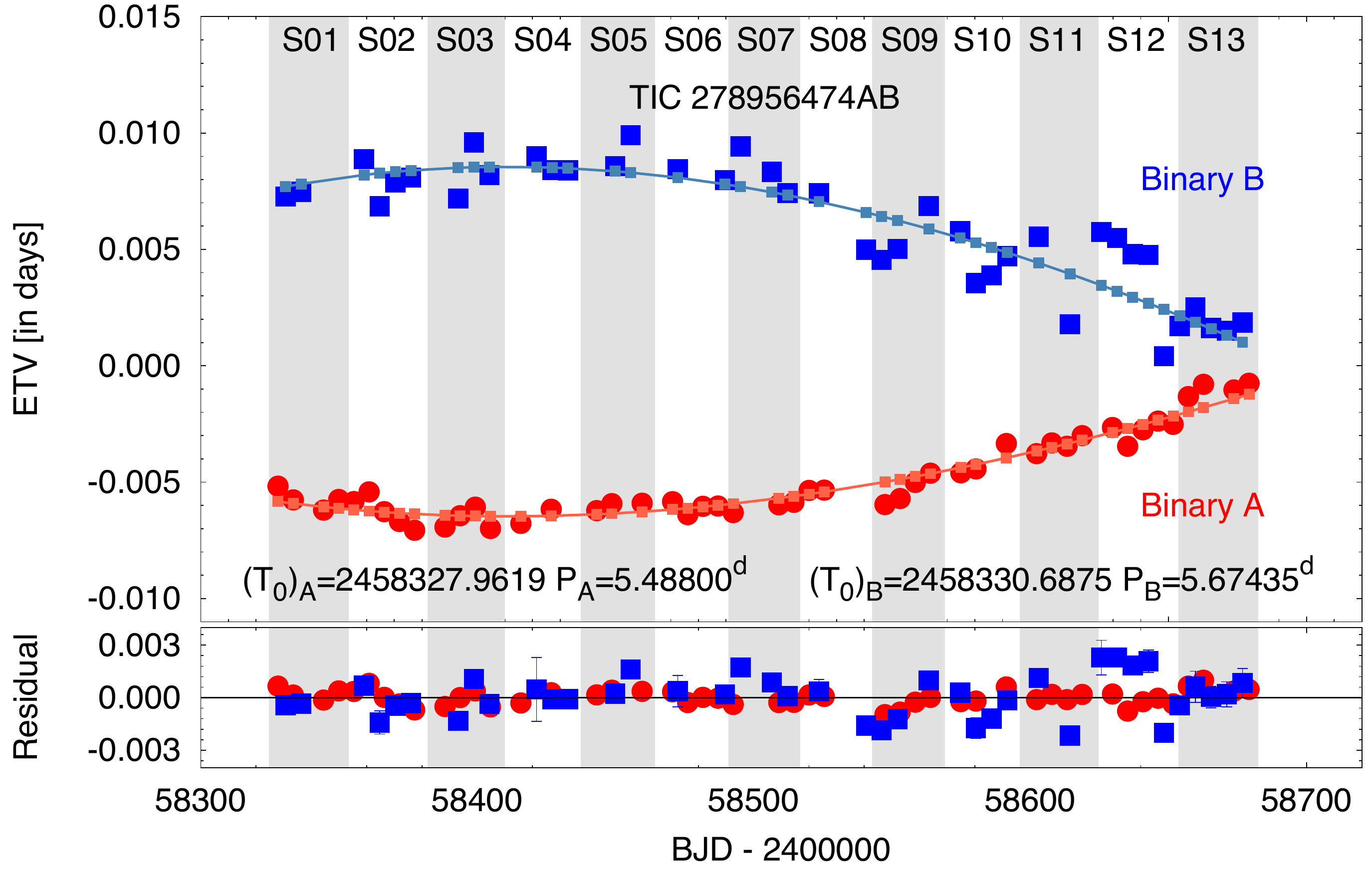}
        \includegraphics[width=0.49 \textwidth]{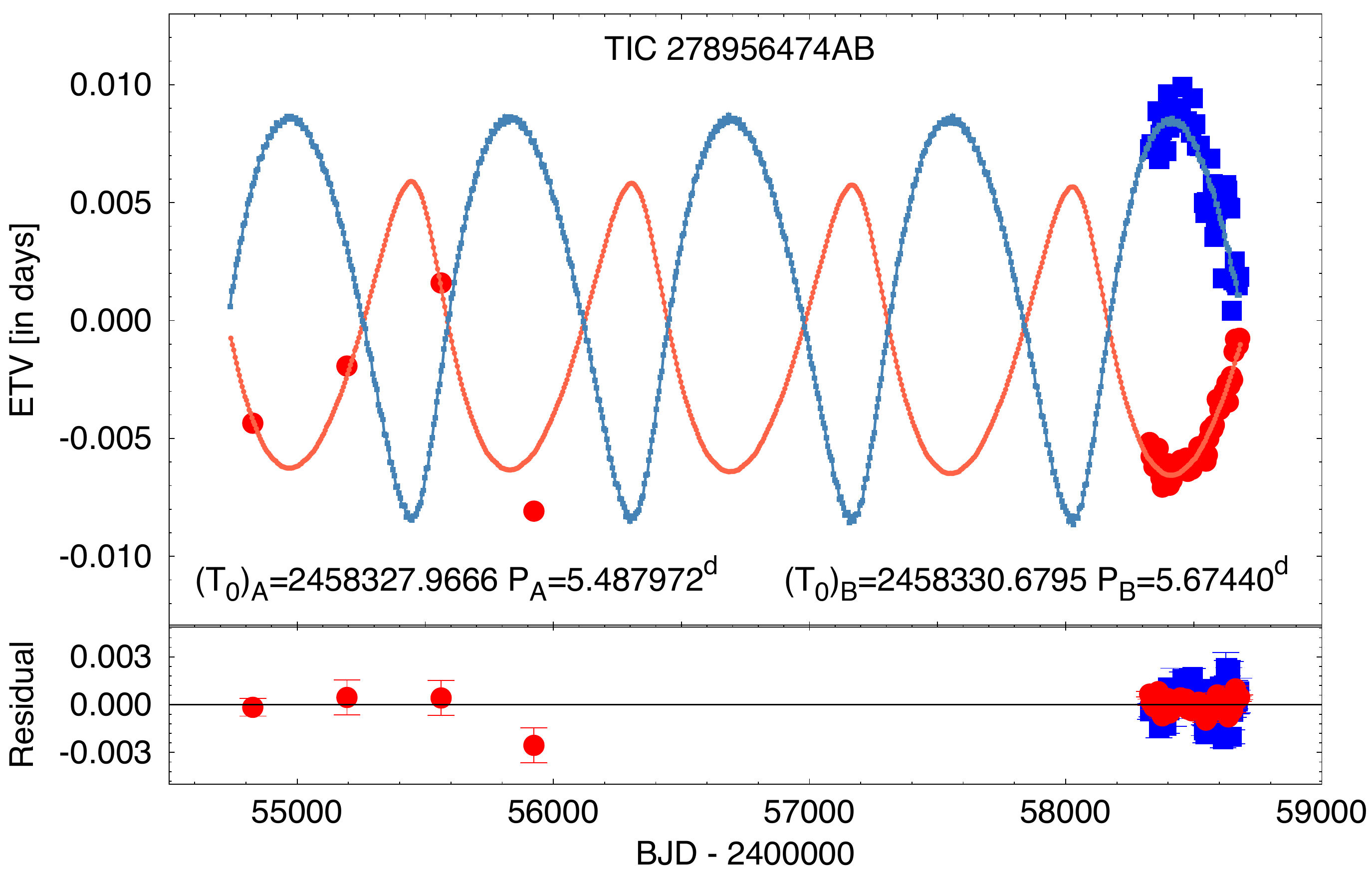}
\caption{Eclipse Timing Variations of the primary minima of TIC 278956474A and B (red and blue, respectively). The anti-correlated nature of the nonlinear timing variations, most likely due to the light-travel time effect, occurs due to the revolution of the barycentres of the two binaries around the common center of mass of the quadruple. Lighter red and blue lines represent the model solutions obtained through the combined light-, ETV and RV-curve analysis, discussed in Section~\ref{lcRVETVanalysis}. Left panel displays the ETVs during the first year of \textit{TESS} observations, whilst the four earlier primary minima of binary A derived from the seasonal average light curves of the historical WASP-South observations are also plotted in the right panel.}
    \label{fig:etv}
\end{figure*}

We plot the ETVs of the primary eclipses of both binaries in the two panels of Fig.~\ref{fig:etv}. The anti-correlated nature of the non-linear timing variations of both binaries are clearly visible. The most likely origin of this feature is the light-travel time effect (LTTE) which arises from the varying distances of the two binaries from the Earth during their revolution around the common center of mass of the whole quadruple system. Therefore, the ETVs strongly suggest that TIC~278956474 is one of the tightest known physically bound 2+2 quadruple systems. This question will be discussed further in Section~\ref{lcRVETVanalysis}. 

\subsection{Gaia DR2}
\label{Gaia}

TIC 278956474 was identified by \textit{Gaia} \citep{2016A&A...595A...1G, 2018A&A...616A...1G, 2018A&A...616A...3R, 2018A&A...616A...8A} as a single source with a mean \textit{Gaia} magnitude of 13.4 and a parallax of $1.08 \pm 0.01$ mas, corresponding to a distance of $926 \pm 12$ pc. Various systematic corrections to \textit{Gaia} parallaxes have been proposed. The online documentation for \textit{Gaia} DR2 states a correction of $-0.03$ mas may be appropriate. The probabilistically derived distance in \citet{2018AJ....156...58B} indicates a distance of $903^{+12}_{-9}$ pc for TIC 278956474, which corresponds to an offset of $-0.03$ mas. \citet{2019MNRAS.487.3568S} finds that on average the parallax offset is $-0.054$ mas; while \citet{2018ApJ...862...61S} find evidence for a systematic offset of $-0.082$ $\pm$ 0.033 mas, for brightnesses G $\ge$ 12 and for distances 0.03-3 kpc. All agree that \textit{Gaia} parallaxes as recorded in the data releases are too small. Despite this, our best fit model (Section~\ref{lcRVETVanalysis}) indicates that the uncorrected \textit{Gaia} parallax for this system is slightly too large and that the true distance is $958\pm23$~pc.

\textit{Gaia} DR2 assigns a Renormalised Unit Weight Error (RUWE) to each source, where a value of 1.0 indicates the source is likely to be a single star, and a value $\ge$ 1.4 indicates that a source is likely to be non-single or otherwise problematic for the astrometric solution, for example a $\le$ 1$''$ binary. The RUWE for TIC 278956474 is 1.06. However, we know from the \textit{TESS} data that TIC 278956474 is a 2+2 quadruple system, not a single star. The low Gaia DR2 RUWE value suggests the two binaries are likely to be tightly bound.

\textit{Gaia} DR2 does not specify an extinction for this system. We use several sources to estimate extinction in the \textit{Gaia} passband. From the catalogue \citet{2019yCat..36250135L}, which uses Gaia and 2MASS photometric data to estimate the extinction toward 27 million carefully selected target stars with a Gaia DR2 parallax uncertainty below 20\%, we estimate the extinction at 0.196 in the \textit{Gaia} passband, although it should be noted that the region in question falls outside the edges of the dust map and hence the reddening is estimated.

Dust maps from \citet{1998ApJ...500..525S} indicate the extinction in the \textit{V} band along the line of sight is 0.198 $\le$ A$_{V}$ $\le$ 0.224 (mean $0.212\pm0.007$), and dust maps from \citet{2011ApJ...737..103S} indicate the extinction along the line of sight is 0.170 $\le$ A$_{V}$ $\le$ 0.193 (mean $0.182\pm0.002$).

An online tool\footnote{\url{https://www.swift.ac.uk/analysis/nhtot/index.php}} estimating $N_{\rm H1}$ and $N_{\rm H2}$ from 493 afterglows detected by the \textit{Swift} X-Ray Telescope \citep{2013MNRAS.431..394W} returns $N_{\rm H,tot}$ $7.76\times10^{20}$ atoms cm$^{-2}$ (mean), $7.34\times10^{20}$ atoms cm$^{-2}$ (weighted). Using the relation between $N_{\rm H}$ and $A_{\rm V}$ in \citet{2009MNRAS.400.2050G}, $A_{\rm V}$ is $0.35\pm0.01$ (mean), $0.33\pm0.01$ (weighted).

This is higher than other estimates, but is in line with the findings in our model of E(B-V) = 0.108$^{+0.025}_{-0.012}$ mag (Section~\ref{lcRVETVanalysis}).

\subsection{Speckle photometry}
\label{Photometry}

\begin{figure}
\centering
	\includegraphics[width=\columnwidth]{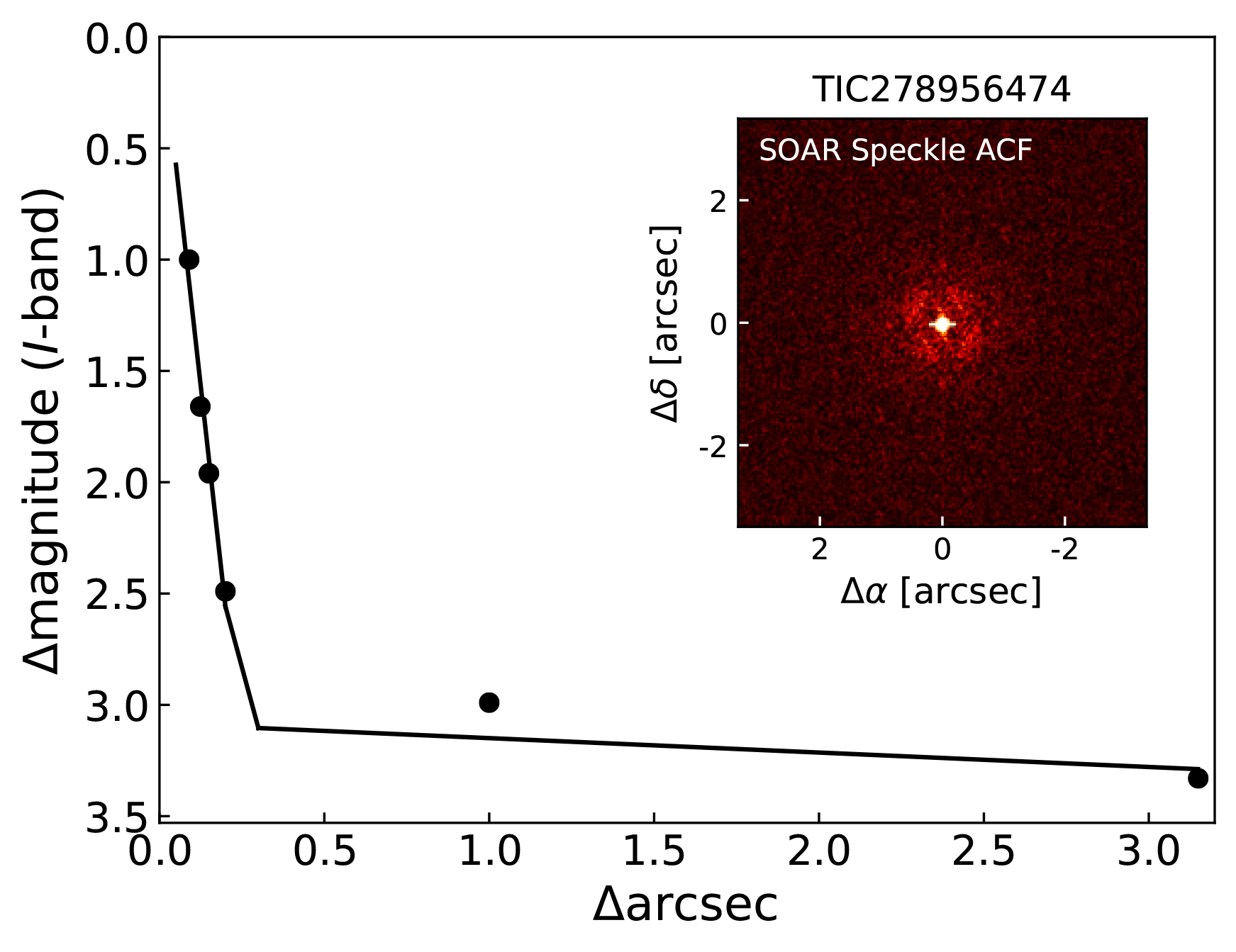}
\caption{Speckle imaging auto-correlation function (inset) and resulting contrast curve obtained on 2020 January 07 with speckle imaging on the 4.1 m Southern Astrophysical Research (SOAR) telescope. This observation places an upper limit for the projected separation of the binaries at approximately 115 AU.}
    \label{fig:soar}
\end{figure}

If the pair of binaries is widely separated, high-angular resolution imaging may be able to resolve the system or detect additional nearby stars. We searched for stellar companions to TIC 278956474 with speckle imaging on the 4.1~m Southern Astrophysical Research (SOAR) telescope \citep{2018PASP..130c5002T} on 2020 January 07 UT, observing in a similar visible bandpass as \textit{TESS}. More details of the observations are available in \citet{2020AJ....159...19Z}. The 5$\sigma$ detection sensitivity and speckle auto-correlation functions from the observations are shown in Fig.~\ref{fig:soar}. The seeing during the night was below average, resulting in a shallow detection curve, and the binaries, assuming a $\Delta_{m}=2$ in the \textit{TESS} bandpass, would likely be resolved at angular separations greater than approximately 0.12$''$, corresponding to a projected separation of $\sim115$ AU at the estimated distance to the system, based on the uncorrected stellar parallax. No nearby stars, however, were detected within 3$''$ of TIC 278956474, placing an upper limit for projected separations of the binaries at approximately 115 AU.

Points that appear a little less than 1$''$ East, West, North, and South of the target are artefacts of the data.

The TESS Input Catalogue (TIC) identifies a 19th magnitude star 11.48$''$ from the target, 106.63$^\circ$ E of N, and a 17th magnitude star 15.77$''$ from the target, 141.72$^\circ$ E of N. Both these stars were also observed by \textit{Gaia}, and \textit{Gaia} does not identify any other stars that are closer. From the difference images in the SPOC data validation reports, it is highly unlikely that the eclipses analyzed here arise from these known near neighbours.

\subsection{Spectroscopy}
\label{Spectroscopy}

\begin{figure}
\centering
	\includegraphics[width=\columnwidth]{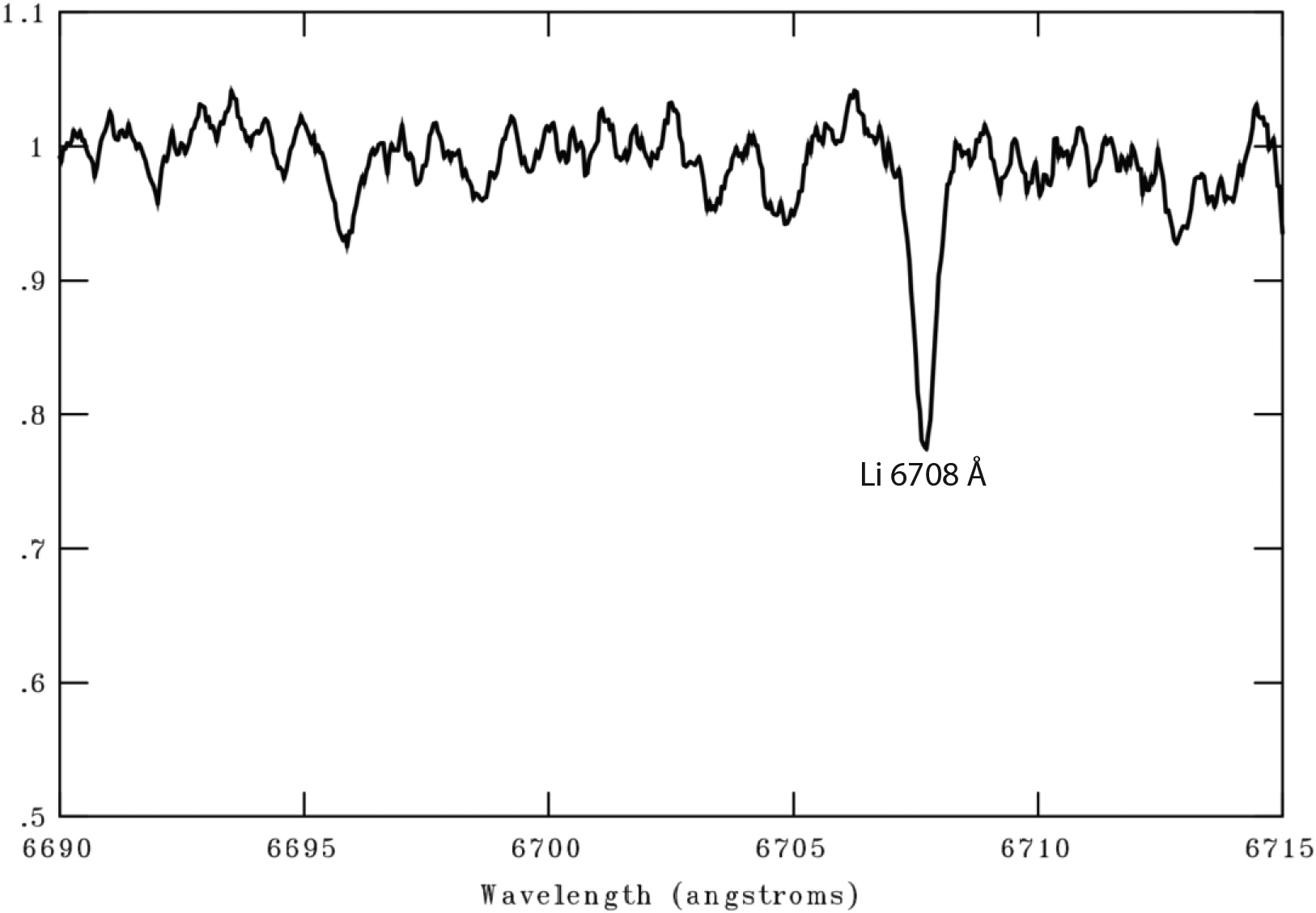}
\caption{Spectra obtained on the High Resolution Spectrograph on the South African Extremely Large Telescope (SALT) indicate a LI 6708 {\AA} equivalent width of 143~$\pm$10~m{\AA} for the whole system.}
    \label{fig:lithium}
\end{figure}

We obtained two spectra using the High Resolution Spectrograph \citep[RSS][]{CrausePerformance2014} on the South African Extremely Large Telescope \citep[SALT;][]{2006SPIE.6267E..0ZB}. We obtained spectra on the nights of 2019 October 03 and 04. The spectra were reduced using the MIDAS pipeline \citep{KniazevMN482016, KniazevSALT2017}\footnote{\url{http://www.saao.ac.za/~akniazev/pub/HRS_MIDAS/HRS_pipeline.pdf}}. The wavelength calibration used ThAr and Ar lamps. The resulting resolution is about $46~000$ and the spectra span 370 to 980 nm.

Separating the components in the spectra is challenging. The second brightest component (Ba) contributes only a few percent of the total light in the optical and any attempt to disentangle the Aa and Ba would be complicated by the presence of the two fainter components, Ab and Bb. We do note absorption of H$\alpha$ and H$\beta$, as well as Ca II at $\approx$ 8664 and 8545~\AA. These features are consistent with our model of the brightest star (Section~\ref{lcRVETVanalysis}).

We also identified a clear and strong Li absorption feature at 6708~\AA~(Fig.~\ref{fig:lithium}). The average equivalent width (EW) is 143~$\pm$10~m\AA. This is almost entirely due to star Aa: the next most significant star, Ba, contributes $\approx$ 3\% of the light.

By comparison with Fig.~4 of \citet{2008AJ....136.2483A}, this EW in a star of the $T_{\rm eff}$ of Aa (6180 K: Section~\ref{lcRVETVanalysis}) is consistent with an age of 30-50 Myr, and by comparison with Fig.~5 of \citet{2008ApJ...689.1127M}, with stars in the $\beta$ Pictoris moving group ($21\pm9$ Myr \citep{2008ApJ...689.1127M}, $22\pm6$ Myr \citep{2017AJ....154...69S}, $24\pm3$ Myr \citep{2015MNRAS.454..593B}) and the Tucanae-Horologium association (isochronal age $20-30$ Myr \citep{2014AJ....147..146K}, $28\pm11$ Myr \citep{2008ApJ...689.1127M}, Li depletion age $\approx 40$ Myr \citep{2014AJ....147..146K}, $45\pm4$ Myr \citep{2015MNRAS.454..593B}).

Estimating $V-K$ for star Aa using the $T_{\rm eff}$, distance and extinction from our model (Section~3.2),  we referred to \citet{2017AJ....153...95R}, which considered the Li depletion (Fig.~21) and ages (Table~1 and references therein) of stars in nearby young moving groups (NYMG). This indicates that TIC 278956474 is likely to be younger than AB Doradus (50-150 Myr), Carina-Near (150-250 Myr) and Ursa Major (300-500 Myr); older than $\epsilon$ Chamaeleontis (5-8 Myr), $\eta$ Chamaeleontis (6-11 Myr) and TW Hydrae (3-15 Myr); and consistent with stars in the following NYMG: $\beta$ Pictoris (10-24 Myr), Octans (20-40 Myr), Tucana-Horologium (30-45 Myr) and Argus (35-50 Myr).

From the Li I EW, we therefore estimate the age of the system to be 10-50 Myr. 

In Fig.~\ref{fig:lithium} the Li 6708$\text{\AA}$ feature has been Doppler shifted to its rest wavelength. Relative to the rest wavelength, the heliocentric RVs are 75.2 $\pm$ 1.8 km s$^{-1}$ on the first night and 36.7 $\pm$ 2.2 km s$^{-1}$ on the second night.

\section{Models}
\label{Models}

\subsection{Preliminary estimates}
\label{tracks}

We made preliminary estimates of the properties of the stars in the system as follows.

Component Ab is fully occluded when it passes behind Aa. \textit{Gaia} obtained 191 astrometric observations of the system, of which 189 were considered good and two bad. Considering that both binaries would have been out of transit for about 86\% of the observing time, and that the difference between the \textit{TESS} magnitude \textit{T} and the \textit{Gaia} magnitude in the red passband $G_{BP}$ is only 0.06 magnitudes and that generally $T \approx G_{BP}$, it is likely that the magnitude of the system in the \textit{G} passband (300-1100 nm) reflects the out-of-transit magnitude of the system. The luminosity of Ab in the \textit{Gaia} passband can therefore be estimated from the total luminosity of the system, taking into account the \textit{Gaia} magnitude, the \textit{Gaia} parallax and an appropriate estimate of extinction (Section~\ref{Gaia}). The luminosity of Ab was estimated as as 0.0247$^{+0.006}_{-0.005}$ L$_\odot$ using the correction from \citet{2018ApJ...862...61S}, and 0.0279$^{+0.008}_{-0.007}$ L$_\odot$ using the uncorrected \textit{Gaia} parallax, in both cases using A$_G$ from \citet{2019yCat..36250135L}.

We used a library of single star evolutionary models from the \texttt{BiSEPS} Binary Stellar Evolution Population Synthesis code \citep{2002MNRAS.337.1004W, 2004A&A...419.1057W, 2006MNRAS.367.1103W, 2010MNRAS.403..179D, 2013MNRAS.433.1133F} to approximate the likely radius, mass, effective temperature and age of Ab, assuming solar metallicity. This allowed us to eliminate the possibility that Ab was a white dwarf.

Comparing the ingress with the total eclipse time during the primary eclipse of A indicated that the ratio of the radii of the two stars was likely to be of the order of 0.29, which indicated that Aa was not evolved and was further confirmation that neither star was a white dwarf. From this and the relationship between the luminosity of Aa and Ab in the \textit{TESS} passband (600-1000 nm centred on 786.5 nm), we estimated the radius and bolometric luminosity of Aa, also at solar metallicity.

We then used stellar evolution tracks from \texttt{MESA} (Modules for Experiments in Stellar Astrophysics) \citep{2011ApJS..192....3P, 2013ApJS..208....4P, 2015ApJS..220...15P} to refine our parameters for Aa and Ab, again assuming solar metallicity. From this we estimated the initial mass and T$_{eff}$ of Aa as 1.315~M$_\odot$ and 6456~K respectively: these estimates would be used as the starting point for the more in depth analysis described in Section~\ref{lcRVETVanalysis}, which makes use of the \texttt{PARSEC} (PAdova and TRieste Stellar Evolution Code) \citep{2012MNRAS.427..127B} stellar evolutionary tracks.

\texttt{BiSEPS} evolves both single stars and binary systems self-consistently from formation to compact remnant. While not a full stellar evolutionary code, the library of models this code produces was useful in obtaining `ballpark' figures for later investigation. \texttt{MESA} and \texttt{PARSEC}, by contrast, are both full stellar evolutionary codes, evolving stars from pre-main sequence to compact remnant. \texttt{MESA} provides information on what is happening in the core as well as on the surface.

\subsection{Combined light, RV and ETV curve analysis with and without joint SED and PARSEC evolutionary track modelling}
\label{lcRVETVanalysis}

We carried out combined, simultaneous analysis of the full {\textit {TESS}} Cycle 1 light curve data together with the ETV data calculated from both \textit{TESS} and WASP-South light curves for the primary eclipses of both binaries (Section\,\ref{ETV}) and also of the two RV data points derived from the spectroscopic observations (Section\,\ref{Spectroscopy}). Several advantages of such a simultaneous analysis are discussed e.~g. in \citet{2018MNRAS.478.5135B}.

For the analysis we prepared the data sets as follows. We downloaded the calibrated two-minute data files for each sector from the MAST  Portal\footnote{\url{https://mast.stsci.edu/portal/Mashup/Clients/Mast/Portal.html}}. For the double binary model analysis we detrended the lightcurve with the software package {\sc W{\={o}}tan} \citep{2019AJ....158..143H}. In this way we removed not only any instrumental effects, but also those light curve variations that might have arisen from the rotation and probable chromospheric activities of the targets, but are not relevant for the binary  star modelling. Then, to save substantial computational time we binned the two-minute cadence data, averaging them every half hour (1800 s). Finally, we kept only those light curve points that were within the $\pm0\fp04$ phase-domain regions around each eclipses. These segments of the light curve were modelled simultaneously with the two ETV curves of the primary eclipses of both binaries (see Section\,\ref{ETV}). Note, some outliers were omitted from the analyzed ETV curves. These points are denoted with an asterix in 
Table\,\ref{Tab:TIC_278956474A_ToM} and Table~\ref{Tab:TIC_278956474B_ToM}.

Finally, we included in the analysis the two RV points (BJD=2\,458\,760.5678; RV$=+75.2\pm1.8\,\mathrm{km s}^{-1}$, and 2\,458\,761.5605; $+36.7\pm2.2\,\mathrm{km s}^{-1}$).

\begin{figure*}
\begin{center}
\includegraphics[width=0.49 \textwidth]{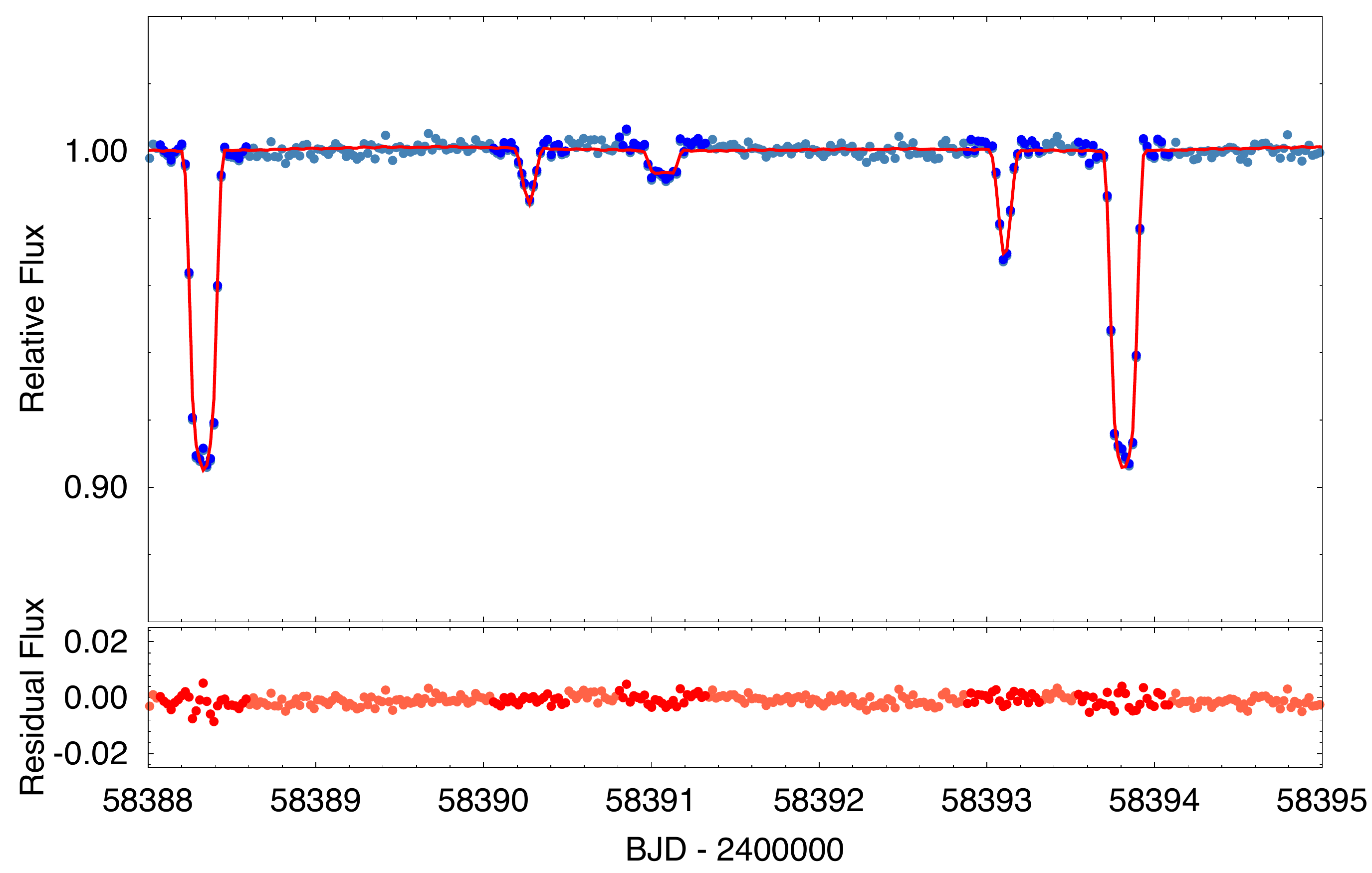}
\includegraphics[width=0.49 \textwidth]{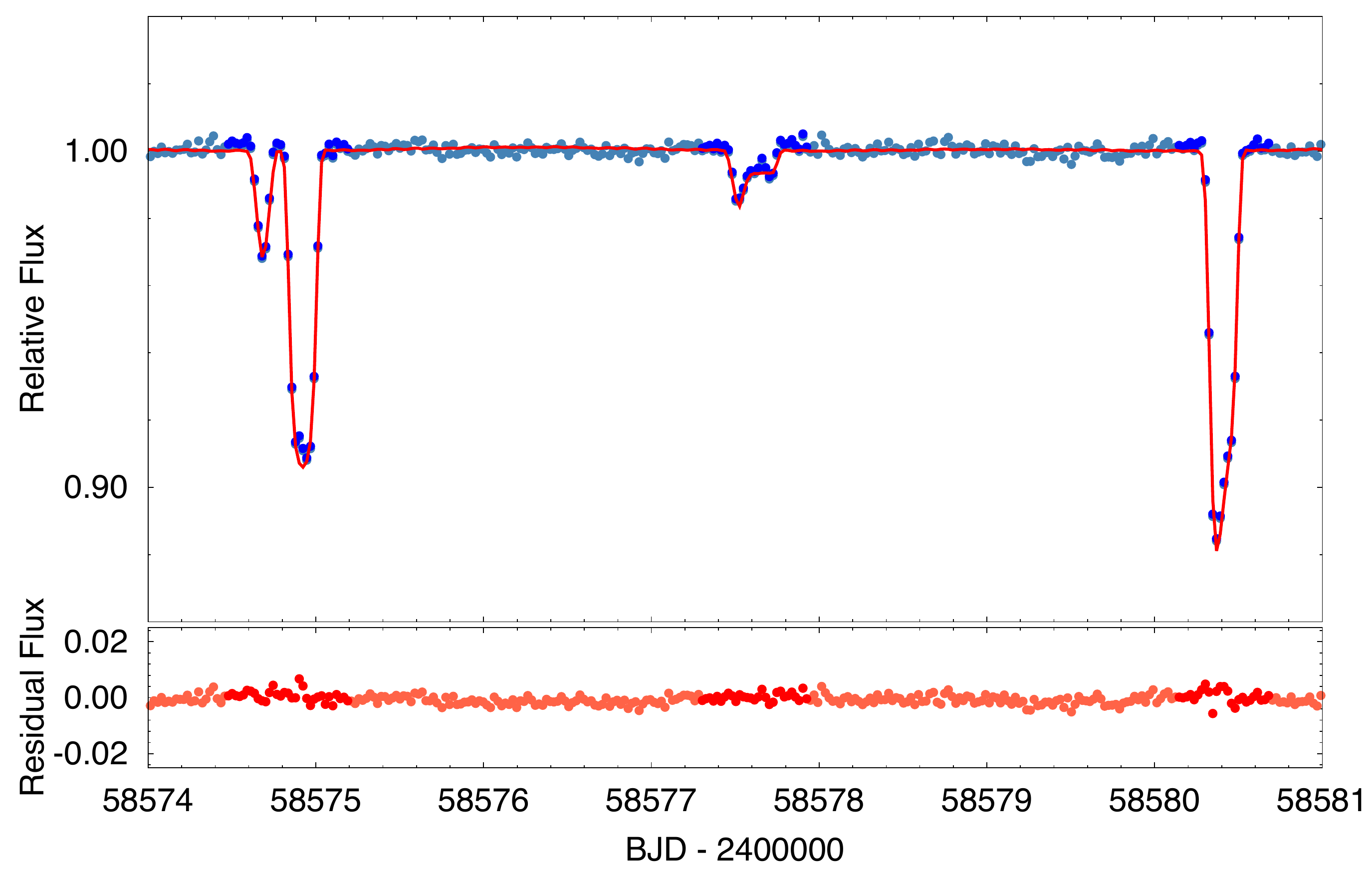}
\caption{Two 7-day-long sections of the \textit{TESS} Year 1 lightcurve of TIC\,278956474. Instead of the full resolution detrended PDCSAP SC flux curve, we plot the 1800-sec binned lightcurve which was used for the photodynamical analysis (see text for details). The dark blue circles in the $\pm0\fp4$ phase-domain around each individual minimum represents the 1800-sec binned flux values used for the photodynamical model, while the other out-of-eclipse data (not used in the modelling) are plotted as light blue circles. The red curve is the cadence-time corrected photodynamical model solution (see Sect.\,\ref{lcRVETVanalysis}); the residuals to the model are also shown in the bottom panels. {\it Left panel:} Here the four different types of eclipses are well separated. From left to right one can see primary eclipse of binary A, secondary of binary B, secondary of binary A, primary of binary B and, finally, the next primary eclipse of binary A. {\it Right panel} illustrates the superpositions of both the secondary (in the middle) and the primary eclipses (to the right) of the two binaries.}
\label{fig:eclipsefitT278} 
\end{center}
\end{figure*}

For our analysis we used software package {\sc Lightcurvefactory} \citep{2018MNRAS.478.5135B,2019MNRAS.487.4631B}. This package is able to model the light- ETV and RV curves of any configurations of eclipsing systems formed by 2--4 stars (i.e. binary, triple and quadruple star systems). For solving the inverse problem, the software employs a Markov chain Monte Carlo (MCMC) parameter search based on an implementation of the generic Metropolis-Hastings algorithm \citep[see e. g.][]{2005AJ....129.1706F}. 

In the first stage of the analysis the temperature ($T_\mathrm{A1}$) and the mass of the primary ($m_\mathrm{A1}$) of binary A were kept fixed on the values given in Section~\ref{tracks}, while the 21 adjusted parameters were as follows:

\begin{itemize}
\item[(i)] 7 light curve related parameters: temperature ratios $(T_2/T_1)_\mathrm{A,B}$ and $T_\mathrm{Ba}/T_\mathrm{Aa}$; the durations of the two primary eclipses $(\Delta t_\mathrm{pri})_\mathrm{A,B}$ (which is closely related to the sum of the fractional radii of the binary stars, see \citealt{2017MNRAS.467.2160R} for an explanation); and the ratios of the radii in both binaries $(R_2/R_1)_\mathrm{A,B})$.
\item[(ii)] $2\times3$ orbital parameters of binaries A and B: we allowed non-zero eccentricities for both binaries and, therefore, $(e\cos\omega)_\mathrm{A,B}$ and $(e\sin\omega)_\mathrm{A,B}$ were freely adjusted. The inclinations of the two orbits ($i_\mathrm{A,B}$) were also adjusted. However, the first sets of runs resulted in insignificantly low inner eccentricites ($e_\mathrm{A,B}\lesssim10^{-3}$), so for later runs we assumed circular inner orbits and, therefore, inner eccentricities and arguments of periastrons were no longer adjusted.
\item[(iii)] 5 orbital parameters of the outer orbit: period ($P_\mathrm{out}$), time of periastron passage $\tau_\mathrm{out}$, eccentricity and argument of periastron as $(e\cos\omega)_\mathrm{out}$, $(e\sin\omega)_\mathrm{out}$ and, finally, the inclination $i_\mathrm{out}$. 
\item[(iv)] 3 mass parameters: the mass ratios of the two binaries ($q_\mathrm{A,B}$) and the mass of the primary of binary B ($m_\mathrm{B1}$).
\end{itemize}

The two periods ($P_\mathrm{A,B}$) and reference primary eclipse times $((T_0)_\mathrm{A,B})$ of both binaries were not adjusted, but constrained through the ETV curves, as was explained in the Appendix A of \citet{2019MNRAS.483.1934B}. Furthermore, the systemic radial velocity of the center of mass of the whole quadruple system ($\gamma$) which in the current model occurs only as an additive parameter independent of any other parameters, was calculated in each trial step by simply minimizing a posteriori the goodness of fit of the RV curve (i.e. $\chi^2_\mathrm{RV}$).

A logarithmic limb darkening law was applied, interpolating the coefficients at each trial step with the use of the pre-computed passband-dependent tables of the {\sc Phoebe} software \citep{2005ApJ...628..426P}.  

Computing the orbital motion and therefore, the sky-projected positions of the four bodies, we assumed purely Keplerian orbits. Though the code has an in-built numerical integrator and therefore, numerical integration of the four-body motion, i.e., application of a photodynamical approach could be done easily, we found it unnecessary for the large $P_\mathrm{out}/P_\mathrm{A,B}$ ratios which render the four-body perturbations undetectable, at least within the time domain of the available observations. 

As a result of this combined analysis we obtained well constrained relative (i.e. dimensionless) stellar parameters (i.e. fractional radii and ratios of temperatures and masses). In order to obtain physical quantities within the frame of a self-consistent model, we added into the analysis the observed cumulative SED of the quadruple, and attempted to find consistent, co-eval \texttt{PARSEC} evolutionary tracks \citep{2012MNRAS.427..127B} for all the four stars. We generated machine readable \texttt{PARSEC} isochrone tables via the web based tool CMD 3.3\footnote{\url{http://stev.oapd.inaf.it/cgi-bin/cmd}}. These tables contain theoretically computed fundamental stellar parameters and absolute passband magnitudes in several different photometric systems, for a large three dimensional grid of ages, metallicities and initial stellar masses. 

\begin{table*}
\centering
\caption{Median values of the parameters from the Double EB simultaneous 
lightcurve and SB1 radial velocity and double ETV and joint SED and PARSEC 
evolutionary tracks solution. $\gamma$ is the systemic radial velocity of the quadruple.}
\begin{tabular}{lccccc}
\hline
\hline
Parameter &
\multicolumn{2}{c}{Binary A} & \multicolumn{2}{c}{Binary B} & Outer orbit \\
\hline
$P$ [days]                       & \multicolumn{2}{c}
{$5.488068_{-0.000010}^{+0.000016}$}   
& \multicolumn{2}{c}{$5.674256_{-0.000030}^{+0.000017}$} & $858_{-5}^{+7}$ 
\\
semimajor axis  [$R_\odot$]      &  \multicolumn{2}{c}
{$15.70_{-0.17}^{+0.09}$}         
& \multicolumn{2}{c}{$14.19_{-0.10}^{+0.11}$} & $543_{-6}^{+5}$ \\  
$i$ [deg]                        & \multicolumn{2}{c}{$88.97_{-0.19}^{+0.16}$}            
& \multicolumn{2}{c}{$89.23_{-0.08}^{+0.16}$} & $85_{-2}^{+3}$  \\
$e$                              &  \multicolumn{2}{c}{$0$}        
& \multicolumn{2}{c}{$0$} & $0.36_{-0.03}^{+0.02}$ \\  
$\omega$ [deg]                   &  \multicolumn{2}{c}{$-$}                
& \multicolumn{2}{c}{$-$} & $299_{-2}^{+2}$ \\
$t_{\rm prim~eclipse}$ [BJD]     & \multicolumn{2}{c}
{$2\,458\,327.9619_{-0.0001}^{+0.0002}$}   
& \multicolumn{2}{c}{$2\,458\,330.6870_{-0.0002}^{+0.0001}$} & 
${2\,458\,930_{-5}^{+5}}^a$\\
$\gamma$ [km/s]                  & \multicolumn{2}{c}{$-$}            & 
\multicolumn{2}{c}{$-$} & $29_{-3}^{+5}$\\  
\hline
individual stars & A1 & A2 & B1 & B2 \\
\hline
Relative Quantities: & \\
\hline
mass ratio [$q=m_2/m_1$]         &\multicolumn{2}{c}
{$0.357_{-0.015}^{+0.009}$}     & 
\multicolumn{2}{c}{$0.876_{-0.049}^{+0.025}$} & $0.691_{-0.016}^{+0.016}$ \\
fractional radius$^b$ [$R/a$]    & $0.1045_{-0.0020}^{+0.0018}$ & 
$0.0306_{-0.0006}^{+0.0005}$  & 
$0.0477_{-0.0023}^{+0.0019}$ & $0.0435_{-0.0014}^{+0.0015}$ \\
fractional luminosity            & $0.927$          & $0.010$           & 
$0.043$          & $0.020$ \\
%extra light  [$l_\mathrm{x}$]    &\multicolumn{4}{c}{$-$} \\
\hline
Physical Quantities: &  \\ 
\hline
$T_{\rm eff}^c$ [K]              & $6180_{-52}^{+99}$  & $3680_{-95}^{+84}$  & 
$4472_{-137}^{+126}$  & 
$3876_{-155}^{+131}$ \\
mass [$M_\odot$]             & $1.271_{-0.046}^{+0.035}$ & 
$0.451_{-0.020}^{+0.016}$ & $0.634_{-0.017}^{+0.022}$ 
& $0.550_{-0.023}^{+0.020}$ \\
radius$^c$ [$R_\odot$]           & $1.641_{-0.046}^{+0.036}$ & 
$0.480_{-0.014}^{+0.011}$ & 
$0.674_{-0.031}^{+0.026}$ & $0.617_{-0.024}^{+0.025}$ \\
luminosity$^c$  [$L_\odot$]          & $3.54_{-0.18}^{+0.20}$ & 
$0.038_{-0.003}^{+0.004}$ & 
$0.16_{-0.01}^{+0.01}$ & $0.079_{-0.012}^{+0.011}$ \\
~~~~~~~~ [$M_\mathrm{bol}$]      & $3.40_{-0.06}^{+0.06}$ & 
$8.32_{-0.10}^{+0.10}$ & 
$6.74_{-0.08}^{+0.10}$ & $7.53_{-0.15}^{+0.18}$ \\
$\log \, g^c$  [cgs]               & $4.11_{-0.01}^{+0.01}$ & 
$4.73_{-0.02}^{+0.02}$ & 
$4.58_{-0.03}^{+0.04}$ & $4.60_{-0.03}^{+0.03}$ \\
$\log$(age) [dex] & $7.00_{-0.05}^{+0.03}$ & $7.90_{-0.07}^{+0.07}$ 
&\multicolumn{2}{c}{$7.70_{-0.11}^{+0.05}$} \\
\hline
$[M/H]$  [dex]                   &\multicolumn{4}{c}{$-0.37_{-0.16}^{+0.10}$} 
\\
$E(B-V)$ [mag]                   &\multicolumn{4}{c}
{$0.108_{-0.012}^{+0.025}$} \\
$(M_V)_\mathrm{tot}^c$             &\multicolumn{4}{c}{$3.39_{-0.06}^{+0.06}$} 
\\
distance [pc]                &\multicolumn{4}{c}{$958_{-23}^{+23}$}\\
\hline
\end{tabular}
\label{tbl:simlightcurve}

{\em Notes.} (a) Time of periastron passage ($\tau_\mathrm{out}$); (b) Polar 
radii; (c) Interpolated from the PARSEC isochrones
\end{table*}

At this final stage of the simultaneous light curve, ETV curves, RV curve, SED and evolutionary track modelling the adjusted parameters have slightly departed from those listed above. First, new adjustable quantities were also introduced, as three independent ages of stars Aa, Ab and binary B\footnote{In the first round we assumed, as usual, that the four stars have the same age but we were unable to find consistent, co-eval solution. Therefore, we decided to allow different stellar ages. This problem will be discussed later.} ($\log\tau_{Aa,Ab,A}$), the metallicity $[M/H]$ of the quadruple, the extinction parameter ($E(B-V)$), and the distance ($d$) of the system. Furthermore, the mass of the most prominent star Aa was no longer fixed, but allowed to adjust with the use of a simple uniform prior. In this way, the actual stellar masses together with the given stellar ages and metallicity, determined the position of each stars on the \texttt{PARSEC} tracks. Then, using a trilinear interpolation with the use of the closest grid points of the precalculated tables, the code interpolated the radii and temperatures of each stars in one hand, and also their absolute passband magnitudes, for the SED fitting, in the other hand. These stellar radii and temperatures were used for the light curve modelling (i.~e., in contrast to the first stage, these quantities were no longer adjusted, but constrained instead). Furthermore, the interpolated absolute passband magnitudes transformed into model observed passband magnitudes  with the use of the extinction parameter and the system's distance, and then, their sum was compared to the observed magnitudes in each passband. In these final steps, distance ($d$) was not a free parameter, but was constrained a posteriori in each trial step by minimizing the value of ($\chi^2_\mathrm{SED}$).

A more detailed description of this joint modelling process, including SED fitting with the use of \texttt{PARSEC} isochrone tables, can be found in Section\,3 of \citet{2020MNRAS.493.5005B}.

The results of this comprehensive analysis are tabulated in Table\,\ref{tbl:simlightcurve} and the model curves are plotted against the observed ETV and light curves in Figs.\,\ref{fig:etv} and \ref{fig:eclipsefitT278}. Moreover, for a better visualisation of the model light curves of both binaries, we also plot the disentangled, phase-folded, binned, averaged versions of the solution light curve against the similarly processed detrended \textit{TESS} light curves of the binaries in Fig.\,\ref{fig:lcfold}. We tabulate the median values of each parameters together with the $1\sigma$ uncertainties. These results will be discussed and compared with an independent SED analysis (Section~\ref{SED}) and \texttt{MESA} evolutionary tracks in Section\,\ref{Discussion}.

\subsection{Independent SED analysis}
\label{SED}

\begin{figure}
\centering
	\includegraphics[width=\columnwidth]{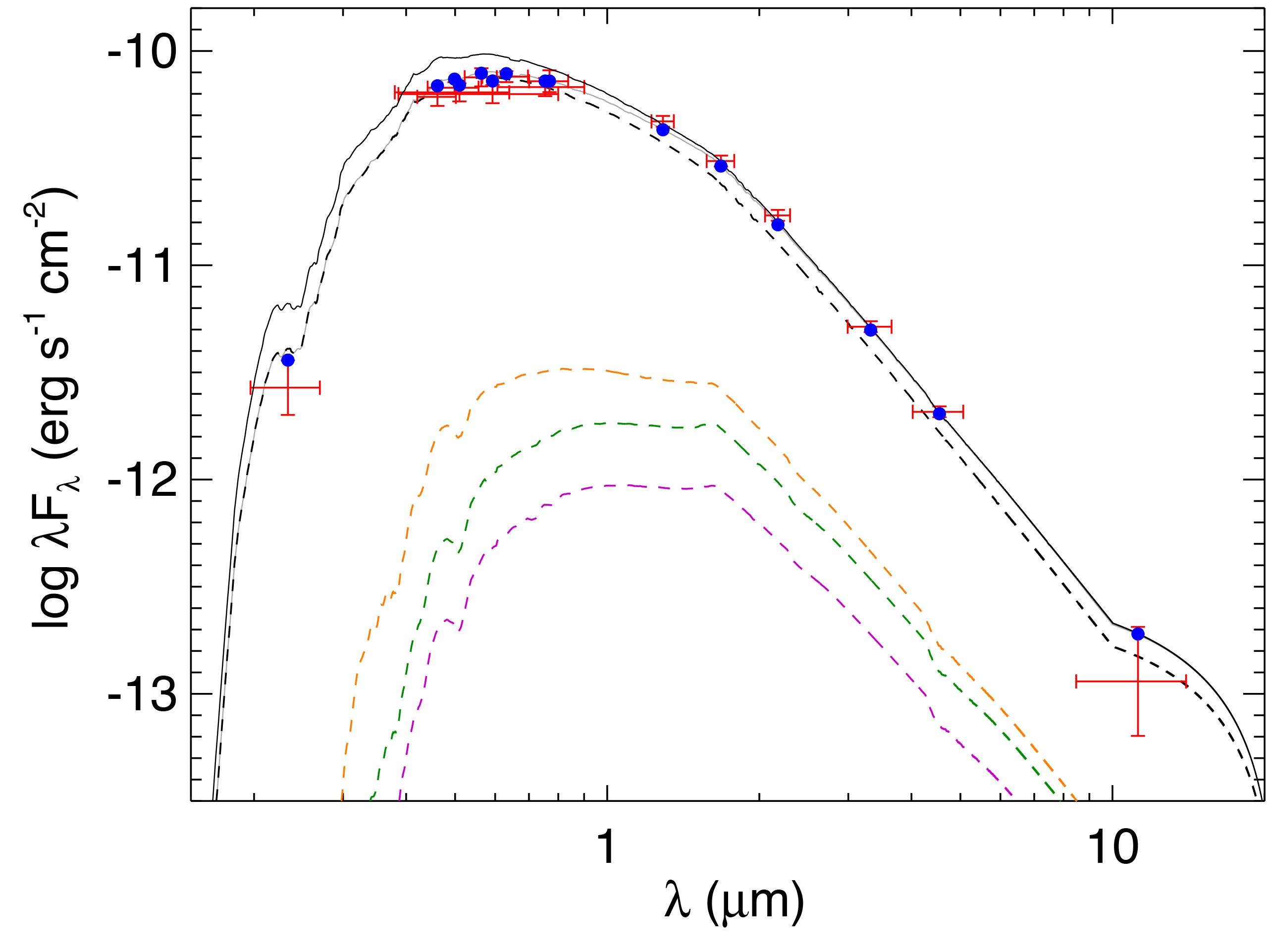}
\caption{Spectral energy distribution (SED). Red symbols represent the observed photometric measurements, where the horizontal bars represent the effective width of the passband. Blue symbols are the model fluxes from the best-fit combined Kurucz atmosphere model (black solid, without extinction; grey, with extinction). Each of the four stellar components is represented by a Kurucz atmosphere of a different color, scaled by the relative stellar surface areas. Black dashed: Aa. Purple: Ab. Orange: Ba. Green: Bb.}
    \label{fig:sed}
\end{figure}

We used the broadband, combined-light spectral energy distribution (SED) of the system, along with the {\it Gaia\/} DR2 parallax, iteratively with the global modeling to check for the possibility of any additional sources of light in the system beyond the four eclipsing components, AaAb+BaBb (Fig.~\ref{fig:sed}). This is separate to the analysis described in Section~\ref{lcRVETVanalysis} and as such provided an independent check on the parameters found. We performed the independent SED modeling with the procedures that \citet{2016AJ....152..180S} developed for EBs, extended here to the case of two EBs simultaneously. 

In brief, a combined-light SED model is calculated from four Kurucz atmospheres \citep{1970SAOSR.309.....K, 2013ascl.soft03024K}, interpolated to the initial estimate values of the individual stellar $T_{\rm eff}$ and scaled by the initial estimate values of the stellar surface areas ($4\pi R_\star^2$). We assume the metallicity identified in the best fit models, [M/H] $-0.37^{+0.10}_{-0.16}$ dex. The remaining free parameters of the fit are the extinction, $A_V$, which we limited to the maximum for the line of sight from the dust maps of \citet{1998ApJ...500..525S}, and the overall flux normalization, $F_{\rm bol,tot}$. We adopted the NUV magnitudes from GALEX (Galaxy Evolution Explorer), $BVgri$ magnitudes from the APASS (The AAVSO Photometric All-Sky Survey) catalog, the $JHK_S$ magnitudes from {\it 2MASS} (Two Micron All Sky Survey), the W1--W4 magnitudes from {\it WISE} (Wide-field Infrared Survey Explorer), and the $G G_{BP} G{RP}$ magnitudes from {\it Gaia}. Together, the available photometry spans the full stellar SED over the wavelength range 0.2--22~$\mu$m (see Fig.~\ref{fig:sed}). 

The resulting fit is excellent (Fig.~\ref{fig:sed}) with a reduced $\chi^2 = 1.2$ and $A_V=0.22^{+0.00}_{-0.02}$ (i.e., the maximum permitted $A_V$ for the line of sight from the dust maps of \citet{1998ApJ...500..525S}, which is expected for the nominal system distance). The total (unextincted) $F_{\rm bol,tot}$ obtained from the observed photometry, together with the model inferred bolometric luminosity ($L_{\rm bol,tot} \equiv 4\pi\sigma_{\rm SB} \Sigma R_\star^2 T_{\rm eff}^4$), yields an implied photometric distance of 964$\pm$13 pc.

This is consistent with the model in Section~\ref{lcRVETVanalysis} ($d_{\rm model} = 958^{+23}_{-23}$~pc) and confirms that the system is likely to be further away than the uncorrected \textit{Gaia} parallax indicates ($d_{\rm Gaia} = 928 \pm 12$~pc). 

\section{Discussion}
\label{Discussion}

\subsection{Examining the model}
\label{Verify}

As demonstrated in Section~\ref{Spectroscopy}, by comparison with other work, the age constraints from the Li I EW are 10-50 Myr. The ages (with uncertainties) of Aa, Ba and Bb in Table~\ref{tbl:simlightcurve} are consistent with this.

The uncertainties in the masses of the four stars may appear to be remarkably low. The combined analysis described in Section~\ref{lcRVETVanalysis} returns mass ratios, fractional radii and fractional luminosity, as indicated in Table~\ref{tbl:simlightcurve}. The mass of one star is required as an input parameter. As discussed in Section~\ref{lcRVETVanalysis}, the mass of Aa was allowed to adjust with the use of a simple uniform prior. The uncertainty in the mass of Aa reflects the `cloud' of solutions which were consistent with the data derived from \textit{TESS} observations, the ETV analysis, the RVs, the SED, data from \textit{Gaia} and the extinction. Stellar evolutionary codes are used to confirm that the age indicated by the physical quantities of the four stars are consistent with the ages indicated by the Li I EW. A wide range of metallicity is indicated, but the stars are clearly sub-solar. 

\begin{figure}
\begin{center}
\includegraphics[width=0.49 \textwidth]{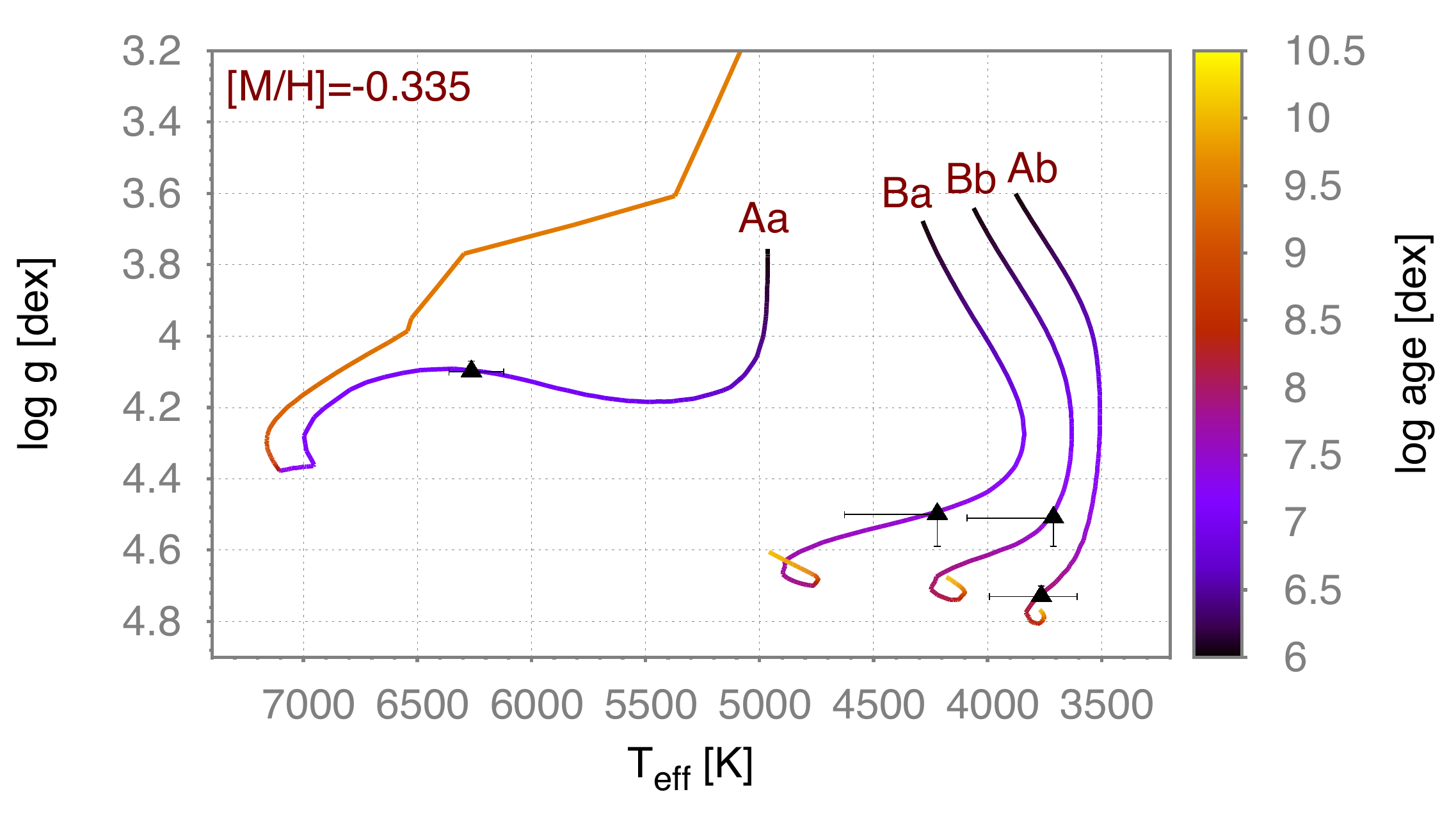}
\caption{$T_\mathrm{eff}$ vs $\log g$ PARSEC evolutionary tracks for the four components of TIC\,27895647 according to the best fitted model. The color scale denotes the age ($\log\tau$) of the stars at any point along their evolution tracks. Black triangles mark the present locations of the four stars in the solution. We note that these positions, probably unphysically, belong to different ages of the given evolutionary tracks.} 
\label{fig:isochrone} 
\end{center}
\end{figure}

\texttt{PARSEC} evolutionary tracks for the four components of TIC\,27895647 from the best fit model are presented in Fig.~\ref{fig:isochrone}. This $T_\mathrm{eff}$ vs $\log g$ plot indicates the position of the best fit models, with uncertainties, with color representing age. As indicated in Table~\ref{tbl:simlightcurve}, the ages are different: Aa is the youngest at 10.0$^{+0.7}_{-1.1}$ Myr, Ab the oldest at 79.4$^{+13.9}_{-11.8}$ Myr, while Ba and Bb have the same age at 50.1$^{+6.2}_{-11.2}$ Myr. Ab is the only star in Table~\ref{tbl:simlightcurve} to have an age which is inconsistent with the Li I EW.

We obtained \texttt{MESA} stellar tracks for the minimum, median and maximum masses indicated by the \texttt{PARSEC} isochrones at Z = 0.01. While this metallicity is not identical to the best fit, it is well within the uncertainty. We match the parameters for Aa at 14$\pm$2 Myr, a good approximation with the \texttt{PARSEC} tracks given the small difference in metallicity. This is consistent with the Li I EW age.

\begin{figure*}
\begin{center}
\includegraphics[width=\textwidth]{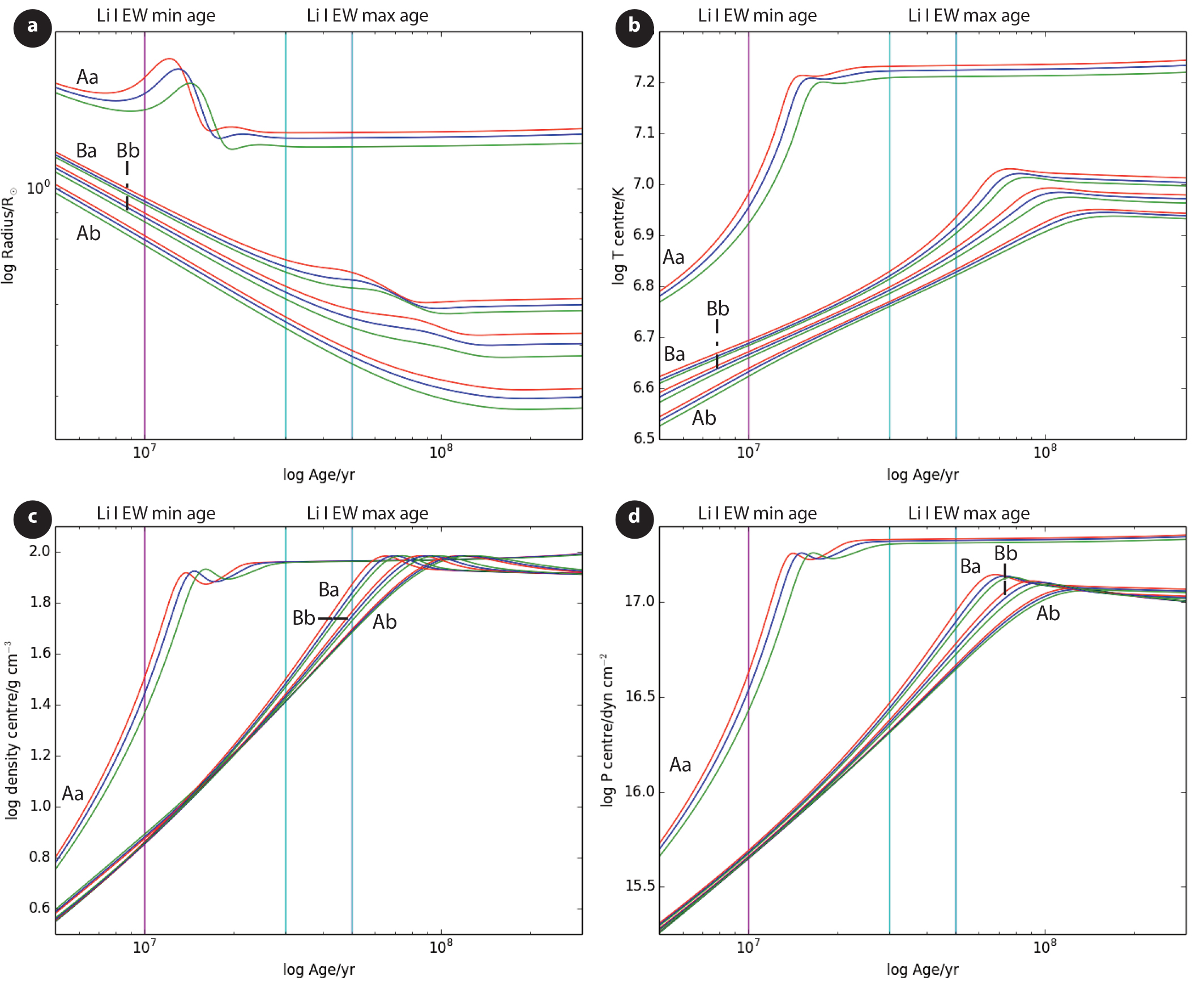}
\caption{\texttt{MESA} isochrones, Z = 0.01, for the median (blue), minimum (green) and maximum (red) masses of the four stars from Table~\ref{tbl:simlightcurve}. The tracks for each star are labelled in each panel. The Li I EW indicates an age between 10-50 Myr, and these limits are also indicated on each panel, along with an age of 30 Myr. Panel a: stellar radius (log scale) from $0.33-2.00~{\rm R}_\odot$; panel b: core temperature; panel c: core density; panel d: core pressure. The point where each star joins the main sequence can be identified from the core characteristics. In these isochrones, which are at a slightly different metallicity to Table~\ref{tbl:simlightcurve} (although within the error bars). the median parameters of Aa are matched at an age of $14\pm2$ Myr, at about the same time as, in these stellar tracks, as Aa joins the main sequence.} 
\label{fig:core} 
\end{center}
\end{figure*}

In Fig.~\ref{fig:core} we compare properties of stars with an initial mass matching the minimum (green), median (blue) and maximum (red) masses from Table~\ref{tbl:simlightcurve}. It appears from panel b (core temperature), panel c (core density) and panel d (core pressure), that $14\pm2$ Myr is approximately the point at which a star of the mass and metallicity of Aa would join the main sequence: in other words, Aa is at ZAMS (zero age main sequence). The feature in panel a (radius), where Aa expands and then contracts at 10-20 Myr, is also evident in the \texttt{PARSEC} isochrones (Fig.~\ref{fig:isochrone}), where $T_\mathrm{eff}$ is plotted against $\log g$. Ab, Ba and Bb would appear to be pre main sequence (PMS) stars.

The model of Ab is the only one which, from Table~\ref{tbl:simlightcurve}, does not have an age consistent with the Li I EW. The ratio of radii of Aa and Ab is strongly constrained by the \textit{TESS} data. From panel a in Fig.~\ref{fig:core}, this points to a lower mass for Ab. Further spectroscopy is required to resolve this issue.

We see no evidence, such as infrared excess in the SED, of a disc in this system, which would appear to be consistent with Aa being at ZAMS rather than still in the T Tauri phase. Our estimate of E(B-V) is consistent with estimates of $N_{\rm H,tot}$ based on \textit{Swift} data (Section~\ref{Gaia}, \citet{2013MNRAS.431..394W}), although higher than other catalogue values.

\subsection{Dynamical properties of the quadruple}
\label{dynamic}

\begin{figure}
\begin{center}
\includegraphics[width=0.49 \textwidth]{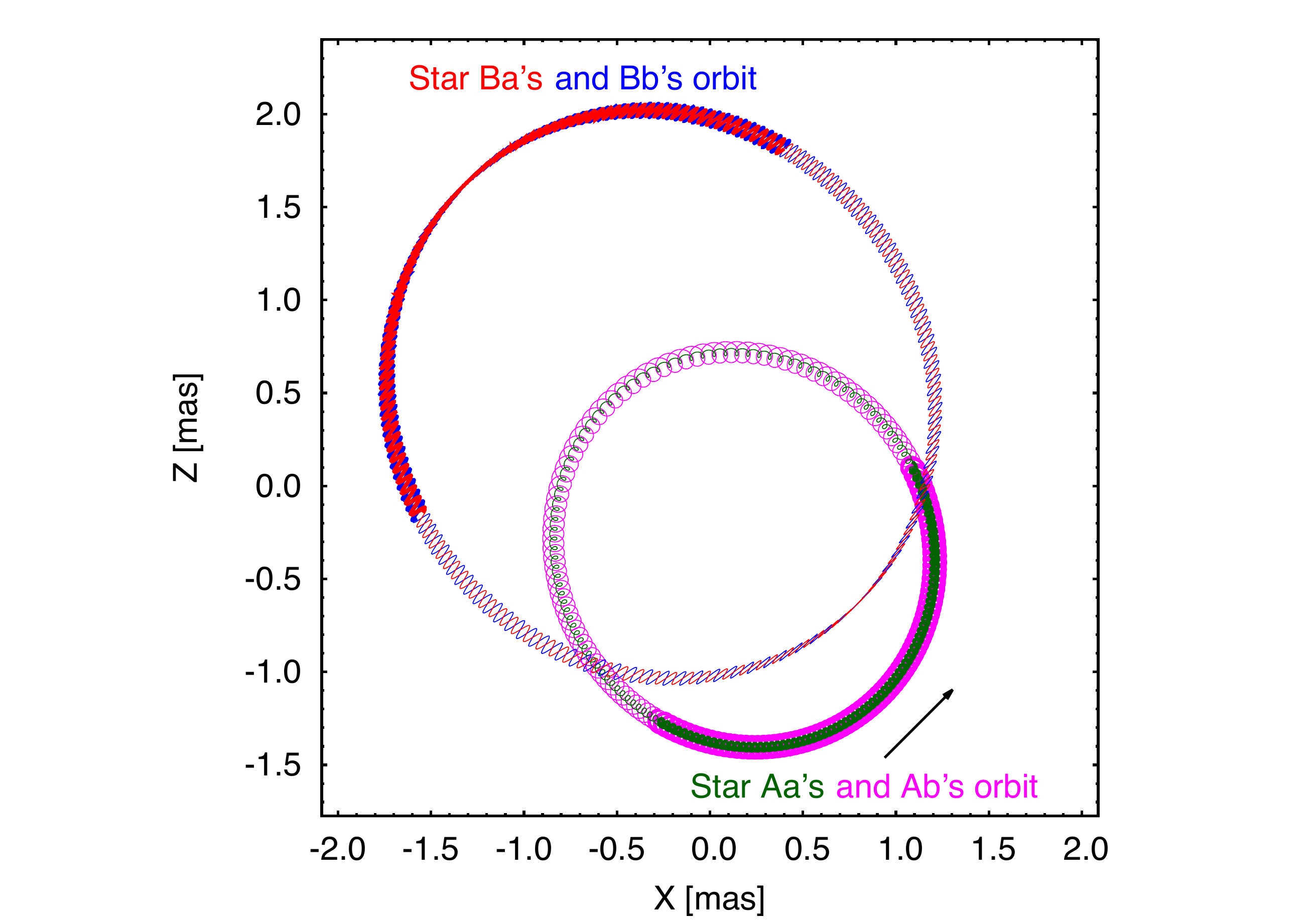}
\caption{The spatial revolutions of the four members of TIC\,278956474, during an outer period, projected to the $X-Z$ plane where the observer is located toward the negative $z$ direction, while the $x$ axis represents the intersection of the orbital plane of the outer orbit with the tangential plane of the sky. The thick arcs represent the four stars' motion during the 11-month-long observations of \textit{TESS} spacecraft. The black arrow shows the direction of the revolution along the outer orbits. The center of mass of the quadruple system is located in the point (0,0). Note, that in the absence of any information about the nodes of the three orbital planes, we assumed quite arbitrarily in this figure that all the three orbital planes intersect the tangential plane of the sky in the same line (i. e. $\Omega_\mathrm{A}=\Omega_\mathrm{B}=\Omega_\mathrm{out}$.)} 
\label{fig:orbconf} 
\end{center}
\end{figure}

From the joint light curve, ETV and RV analysis TIC\,278956474 was found to be one of the most compact known 2+2 quadruple stellar systems. We display the spatial configuration of the system in Fig.\,\ref{fig:orbconf}. The median period of the outer orbit was found to be $P_\mathrm{out}=858_{-5}^{+7}$\,d, with a moderate eccentricity of $e_\mathrm{out}=0.36_{-0.03}^{+0.02}$. (For comparison, note that the tightest known 2+2 quadruple system, VW~LMi, has an outer period of $P_\mathrm{out}=355$\,d and eccentricity of $e_\mathrm{out}<0.1$, see \citealp{2008MNRAS.390..798P,2020MNRAS.494..178P}.) One should keep in mind, however, that while the bound quadruple nature of the system is certainly beyond question, the quantitative results on the orbital parameters of the outer orbit should be considered only with caution. The reason is, that the \textit{TESS} observations cover only a fraction of an outer orbital period and, furthermore, the four former, seasonal WASP minima have large uncertainties. Furthermore, as one can see in Fig.\,\ref{fig:orbconf}, the present solution suggests that the system was observed around the apastron phase of the outer orbit, i.e. when the orbital motion is the slowest and, therefore, the curvatures of ETV curves are also minimized. Therefore, future follow up eclipse timing observations would be extremely useful to obtain more certain outer orbital parameters.

In contrast to the outer orbit, the obtained elements of the two close binary orbits should be robust. The period ratio of the two binary orbits ($P_\mathrm{B}/P_\mathrm{A}\sim1.03$) is very close to unity. According to the results of \citet{2019A&A...630A.128Z} there is a significant excess of 2+2 quadruple systems with near equal inner periods, however, the origin of this feature is still unknown \citep[see also][]{2018MNRAS.475.5215B,2020MNRAS.493.5583T}. Turning to the other orbital parameters, as preliminary runs implied that eccentricities of both inner orbits should be less than 0.001, we assumed circular orbits for the further analysis. This assumption does not contradict the young age of the system, as it was shown by \citet{1989A&A...223..112Z} that the orbits of close binaries formed by late type stars and having period $P\lesssim7-8$\,d, are expected to circularise by the end of the very first one million years of their pre-MS evolution.

Regarding the binary-binary mutual (gravitational) interactions, their period and amplitude can be estimated from the analytical theory of the perturbations in hierarchical triple systems \citep[see, e.\,g.][]{1975A&A....42..229S}. The key parameter is the period ratio of $P_\mathrm{out}/P_\mathrm{in}$ which is $\approx150$ for both binaries. According to the results of \citet{2015MNRAS.448..946B,2016MNRAS.455.4136B} for such a high value, the short term perturbations are negligible and, therefore, within the time-scale of the $\sim1$-yr-long \textit{TESS} observations, the orbital motion of the four stars along their inner and outer orbits can be considered as unperturbed Keplerian motions. As a consequence, neither the \textit{TESS} photometry, nor the RV measurements carry any information about the nodal angles of the three orbital planes and, therefore, despite the fact that the angles of inclination of A ($i_\mathrm{A}=88.97^{+0.16}_{-0.19}$ deg) and B ($i_\mathrm{B}=89.23^{+0.16}_{-0.08}$ deg), and also of the outer orbit ($i_\mathrm{out}=85^{+3}_{-2}$ deg) were found to be very similar, one cannot declare that the orbits are almost coplanar. Oppositely, one can say nothing about the mutual inclinations of any two of three orbital planes.\footnote{Strictly speaking what we can obtain uniquely is only $\sin i$, therefore, even these three inclination values are ambiguous for the undetermined signs of $\cos i$-s.}

Considering the larger amplitude, so-called apse-node timescale (or, secular) perturbations, their characteristic period is proportional to $P_\mathrm{out}^2/P_\mathrm{in}\approx350$\,yr for our quadruple. In such a way, in the case of a non-coplanar configuration, we can expect significant inclination and therefore, well-observable eclipse depth variations within a few decades.

\subsection{Young quadruple systems}

From the Li I EW (Section~\ref{Spectroscopy}), this quadruple system is young (10-50 Myr). In the introduction we referred to \citet{2012A&A...543A...8M}, who demonstrated through simulations that if a cluster started with a formal binary fraction of unity, the binary fraction would reduce over time to match presently observed values. Our analysis of this system indicates that even small groupings of stars may form as hierarchical multiples. Whether systems such as TIC 278956474 are likely to become unbound over time would be an interesting follow-up study.

Of the young quadruple systems mentioned in Section~\ref{Intro}, some are young enough to be consistent with T Tauri stars. GG Tauri and HD 98800, for example, have evidence of circumbinary discs around one of the inner binaries \citep{1999A&A...348..570G, 2018ApJ...865...77R, 2000ApJ...533L..37K}, as well as periods significantly longer than that of TIC 278956474. The inner binaries in GG Tauri have periods of $\approx$ 403 days and 40,000 years \citep{2011A&A...530A.126K}, while the inner binaries of HD 98800 have periods of about 315 days (\citet{2018ApJ...865...77R} and references therein). The inner period of HD 34700 has been measured at 23.5 days \citep{2004AJ....127.1187T, 2005A&A...434..671S}. Again, there is evidence of a circumstellar disc \citep{2015ApJ...809...22S}. We have no evidence of a circumstellar disc around TIC 278956474, nor would we expect to find any in a system where the brightest star is at ZAMS.

LkCa 3 is a young quadruple system, with an age of about 1.4 Myr \citep{2013ApJ...773...40T} where all four components are M-class stars. This is interesting as it is generally accepted that high mass stars are more likely to be binary than are low mass stars. Our final model of TIC 278956474 suggests while the spectral classes of the stars will eventually be F3 (Aa), M0 (Ab), K3 (Ba) and K7 (Bb), presently both Ab and Bb could be considered M-class. 

\citet{2006A&A...449..327G} report the age of AO Vel to be ZAMS, with two short period inner binaries: 1.58 days and 4.15 days. The outer period is reported as 41.0 years. This system is listed on Simbad as an Algol-type eclipsing binary. The inner periods of AO Vel are shorter than those of TIC 278956474, but do not share the characteristic of being very similar in duration.

\citet{2014A&A...570A..95W} revisited the age of the quadruple system AB Doradus (as opposed to the moving group of the same name), and concludes that it is 50-100 Myr old. This is older than TIC 278956474 and the periods are longer. One inner binary has a period of 11.7 years, the other has a period of 361 days, and the outer period is estimated at $\approx$ 1500 yr \citep{2014A&A...570A..95W}.

HD 91962 is unusual in that three companions appear to orbit one central star \citep{2015AJ....149..195T}. The system is considered to be young on the basis of lithium abundances, and the three periods are 170.3 days, 8.85 years and 205 years. \citet{2015AJ....149..195T} put forward the theory that this system was formed when companions migrated in a dissipative disc formed from the collapse of an isolated core. TIC 278956474 may also have formed from the collapse of an isolated core, but has a more conventional architecture.

HD 86588 is dated to 10 $\lesssim$ Myr $\lesssim$ 150 \citep{2018AJ....156..120T} and therefore overlaps in terms of age with TIC 278956474, although not in architecture. The four stars, with masses 1 $\lesssim$ M/M$_\odot$ $\lesssim$ 1.3, are in a three-tier hierarchy. The inner period, 2.4058 days, has not yet fully circularised as the eccentricity is 0.086 $\pm$ 0.003. By contrast, in TIC 278956474 the two inner periods have circularised. \citet{2018AJ....156..120T} state the outermost period of HD 86588 is around 300 years, and that the intermediate period is 8 years.

Young quadruple systems come in many guises, and TIC 278956474 adds to our understanding of such systems. Both inner periods are known to be short, are circularised and are similar in duration to each other. HD 34700 also has a short inner period. AO Vel also has two short period inner binaries, but one inner period is $\approx$~2.6 times as long as the other. The inner period of HD 86588 is shorter than in TIC 278956474 but has not yet circularised: both inner periods in TIC 278956474 are already circular. The other young quadruple systems tend towards longer inner periods.

In terms of architecture, as a 2+2 quadruple system TIC 278956474 is similar to most other known young quadruple systems. Other architectures are observed, but appear to be less common.

Of the eight systems other than TIC 278956474 discussed here, HD 98800 appears to have no entry in \textit{Gaia} DR2, but has a known distance of 47 pc \citep{1998ApJ...498..385S}. LkCa3 has no parallax recorded in \textit{Gaia} DR2 but has a known distance of 133 pc \citep{2013ApJ...773...40T}. GG Tauri has a negative parallax in \textit{Gaia} DR2, but has a known distance of 140 pc (\citet{2019A&A...628A..88B} and references therein). The remaining systems all have reliable parallaxes in \textit{Gaia} DR2. AB Doradus ($65.3 \pm 0.1$ mas), HD 91962 ($28.2 \pm 0.5$ mas) and HD 34700 ($2.81 \pm 0.05$ mas) all have parallaxes indicating they are closer to us than is TIC 278956474. AO Vel ($1.12 \pm 0.04$ mas) and HD 86588 ($1.00 \pm 0.05$ mas) have a similar parallax to, and hence are at about the same distance as, TIC 278956474. None appear to be significantly further away.

Because of its magnitude and distance, without \textit{TESS} observations and the SPOC pipeline processing, TIC 278956474's nature as a quadruple system, rather than a single star, would not have been identified.

\section{Conclusion}
\label{Conclusion}

The \textit{TESS} mission has enabled the identification of a TIC 278956474 as a 2+2 quadruple system composed of two short-period eclipsing binaries, where all previous observations, including \textit{Gaia}, indicated that this system was one single star. 

The eclipses detected in the SPOC pipeline cannot be planetary in nature. Aa and Ab have the same period as each other, as do Ba and Bb. The periods of Aa+Ab and of Ba+Bb, 5.488 days and 5.674 days respectively, would not appear to be consistent with a stable planetary system. While highly inflated short period planets can have secondaries, the eclipses of both Ab and Bb are too deep for this to be a reasonable explanation.

Using SPOC data validation reports, archival WASP-South data, a study of eclipse timing variations, speckle photometry, spectroscopy, data from \textit{Gaia} DR2 and the \texttt{BiSEPS}, \texttt{MESA} and \texttt{PARSEC} stellar evolutionary codes, we identified a model for the four stars in this system which is consistent with observations, the SED and the uncorrected \textit{Gaia} parallax. It is unlikely that there are additional detectable components. The best fit parameters are given in Table~\ref{tbl:simlightcurve}.

The Li I EW indicates an age of 10-50 Myr, and our model is consistent with this approximation. One star appears to be at ZAMS, while the other three are still on the PMS. 

Further observations, in particular spectroscopy and photometry, would be valuable in refining the properties of this system, in particular the parameters of the outer period (Fig.~\ref{fig:orbconf}).

\section*{Acknowledgements}
The authors thank the anonymous referee for his or her helpful comments.

This paper includes data collected by the \textit{TESS} mission, which are publicly available from the Mikulski Archive for Space Telescopes (MAST). Funding for the \textit{TESS} mission is provided by NASA's Science Mission directorate.

Resources supporting this work were provided by the NASA High-End Computing (HEC) Program through the NASA Advanced Supercomputing (NAS) Division at Ames Research Center for the production of the SPOC data products.

This work has made use of data from the European Space Agency (ESA) mission {\it Gaia} (\url{https://www.cosmos.esa.int/gaia}), processed by the {\it Gaia} Data Processing and Analysis Consortium (DPAC, \url{https://www.cosmos.esa.int/web/gaia/dpac/consortium}). Funding for the DPAC has been provided by national institutions, in particular the institutions participating in the {\it Gaia} Multilateral Agreement.

T.\,B. acknowledges the financial support of the Hungarian National Research, 
Development and Innovation Office -- NKFIH Grant KH-130372.

The authors also thank Oleg Khozhura for his contribution to this paper.

\software{BiSEPS \citep{2002MNRAS.337.1004W, 2004A&A...419.1057W, 2006MNRAS.367.1103W, 2010MNRAS.403..179D, 2013MNRAS.433.1133F}, gnuplot (\url{http://www.gnuplot.info/}), Lightcurvefactory \citep{2018MNRAS.478.5135B, 2019MNRAS.487.4631B}, MESA \citep{2011ApJS..192....3P, 2013ApJS..208....4P, 2015ApJS..220...15P}, PARSEC \citep{2012MNRAS.427..127B}, W{\={o}}tan \citep{2019AJ....158..143H}.}

\bibliographystyle{aasjournal}
\bibliography{main2}

\appendix{SPOC data}
\begin{table*}
\caption{Data from SPOC. The secondary in component Aa coincides with the time of the primary in component Ab, and vice versa. Similarly for components Ba and Bb. Secondary depths are therefore only included where each component is not separately identified. A second science run was completed for multisector 1-13 in order to identify component Ab at the correct period and to remove the partial eclipse of Aa at the end of sector 8. This science run is identified as sector 1-13*.}
\centering
\label{table:SPOC}
\scalebox{0.65}{
\begin{tabular}{l l l l l l l l l}
\toprule
Component & Sector & Period/days & Depth/ppm & Duration/hr & Ingress/hr & Odd depth/ppm & Even depth/ppm & Secondary depth/ppm\\
\midrule
Aa & 1 & 5.4881 $\pm$ 0.0002 & 94800 $\pm$ 900 & 5.50 $\pm$ 0.04 & 1.22 $\pm$ 0.04 & 94000 $\pm$ 1000 & 96000 $\pm$ 1000 & Component Ab\\
Aa & 1-2 & 5.48797 $\pm$ 0.00008 & 94600 $\pm$ 600 & 5.44 $\pm$ 0.03 & 1.21 $\pm$ 0.02 & 95100 $\pm$ 900 & 94500 $\pm$ 800 & Component Ab\\
Aa & 1-3 & 5.48789 $\pm$ 0.00004 & 94200 $\pm$ 500 & 5.43 $\pm$ 0.02 & 1.21 $\pm$ 0.02 & 93900 $\pm$ 600 & 94500 $\pm$ 600 & Component Ab\\
Aa & 1-6 & 5.48797 $\pm$ 0.00002 & 93900 $\pm$ 300 & 5.41 $\pm$ 0.01 & 1.20 $\pm$ 0.01 & 95700 $\pm$ 500 & 92900 $\pm$ 500 & Component Ab\\
Aa & 1-9 & 5.487995 $\pm$ 0.000008 & 94200 $\pm$ 300 & 5.43 $\pm$ 0.01 & 1.21 $\pm$ 0.01 & 94400 $\pm$ 400 & 94000 $\pm$ 400 & Component Ab\\
Aa & 1-13 & 5.488035 $\pm$ 0.000005 & 93900 $\pm$ 200 & 5.43 $\pm$ 0.01 & 1.22 $\pm$ 0.01 & 94600 $\pm$ 300 & 93300 $\pm$ 300 & 8900 $\pm$ 400\\
Aa & 1-13* & 5.488036 $\pm$ 0.000005 & 93900 $\pm$ 200 & 5.43 $\pm$ 0.01 & 1.22 $\pm$ 0.01 & 94500 $\pm$ 300 & 93400 $\pm$ 300 & Component Ab\\
Aa & 2 & 5.4878 $\pm$ 0.0002 & 94700 $\pm$ 900 & 5.42 $\pm$ 0.03 & 1.22 $\pm$ 0.03 & 94000 $\pm$ 1000 & 97000 $\pm$ 2000 & Component Ab\\
Aa & 3 & 5.4882 $\pm$ 0.0003 & 93700 $\pm$ 600 & 5.43 $\pm$ 0.03 & 1.20 $\pm$ 0.03 & 94300 $\pm$ 800 & 92400 $\pm$ 900 & Component Ab\\
Aa & 4 & 5.4882 $\pm$ 0.0003 & 93400 $\pm$ 800 & 5.34 $\pm$ 0.04 & 1.18 $\pm$ 0.04 & 98000 $\pm$ 1000 & 91000 $\pm$ 1000 & Component Ab\\
Aa & 5 & 5.4884 $\pm$ 0.0003 & 94300 $\pm$ 800 & 5.47 $\pm$ 0.03 & 1.24 $\pm$ 0.03 & 94000 $\pm$ 1000 & 95000 $\pm$ 1000 & Component Ab\\
Aa & 6 & 5.4881 $\pm$ 0.0003 & 95700 $\pm$ 800 & 5.41 $\pm$ 0.03 & 1.21 $\pm$ 0.03 & 96000 $\pm$ 1000 & 96000 $\pm$ 1000 & Component Ab\\
Aa & 7 & 5.4877 $\pm$ 0.0002 & 93000 $\pm$ 1000 & 5.38 $\pm$ 0.04 & 1.21 $\pm$ 0.03 & 88000 $\pm$ 2000 & 97000 $\pm$ 2000 & 7000 $\pm$ 2000\\
Aa & 8 & 5.4873 $\pm$ 0.0002 & 93000 $\pm$ 1000 & 5.46 $\pm$ 0.03 & 1.21 $\pm$ 0.03 & 56378 $\pm$ 1000 & 95147 $\pm$ 2000 & Component Ab\\
Aa & 9 & 5.4888 $\pm$ 0.0003 & 93400 $\pm$ 900 & 5.45 $\pm$ 0.03 & 1.27 $\pm$ 0.03 & 94000 $\pm$ 1000 & 93293 $\pm$ 1000 & Component Ab\\
Aa & 10 & 5.4890 $\pm$ 0.0003 & 93600 $\pm$ 700 & 5.53 $\pm$ 0.04 & 1.31 $\pm$ 0.04 & 98000 $\pm$ 1000 & 90000 $\pm$ 1000 & Component Ab\\
Aa & 11 & 5.4885 $\pm$ 0.0003 & 92100 $\pm$ 800 & 5.41 $\pm$ 0.04 & 1.20 $\pm$ 0.04 & 91000 $\pm$ 1000 & 93000 $\pm$ 1000 & Component Ab\\
Aa & 12 & 5.4880 $\pm$ 0.0003 & 95000 $\pm$ 1000 & 5.42 $\pm$ 0.04 & 1.2 $\pm$ 0.04 & 95000 $\pm$ 2000 & 94000 $\pm$ 2000 & Component Ab\\
Aa & 13 & 5.4882 $\pm$ 0.0003 & 94000 $\pm$ 1000 & 5.43 $\pm$ 0.04 & 1.21 $\pm$ 0.04 & 94000 $\pm$ 2000 & 94000 $\pm$ 2000 & 8000 $\pm$ 2000\\
\midrule
Ab & 1 & 5.495 $\pm$ 0.003 & 8500 $\pm$ 800 & 5.5 $\pm$ 0.4 & 1.1 $\pm$ 0.4 & 8000 $\pm$ 30000 & 8000 $\pm$ 1000 & Component Aa\\
Ab & 1-2 & 5.4883 $\pm$ 0.0008 & 9100 $\pm$ 600 & 5.3 $\pm$ 0.2 & 1.1 $\pm$ 0.2 & 8900 $\pm$ 800 & 8300 $\pm$ 800 & Component Aa\\
Ab & 1-3 & 5.4881 $\pm$ 0.0004 & 9100 $\pm$ 400 & 5.3 $\pm$ 0.2 & 1.0 $\pm$ 0.2 & 9400 $\pm$ 500 & 8200 $\pm$ 600 & Component Aa\\
Ab & 1-6 & 5.4882 $\pm$ 0.0001 & 9000 $\pm$ 300 & 5.3 $\pm$ 0.1 & 0.9 $\pm$ 0.1 & 9300 $\pm$ 400 & 8400 $\pm$ 400 & Component Aa\\
Ab & 1-9 & 5.48809 $\pm$ 0.00007 & 8900 $\pm$ 200 & 5.4 $\pm$ 0.1 & 1.1 $\pm$ 0.1 & 9200 $\pm$ 300 & 8600 $\pm$ 400 & Component Aa\\
Ab & 1-13 & 2.74403 $\pm$ 0.00002 & 8900 $\pm$ 200 & 5.29 $\pm$ 0.09 & 1.0 $\pm$ 0.1 & Model fitter failed & Model fitter failed & n/a\\
Ab & 1-13* & 5.48808 $\pm$ 0.00004 & 8900 $\pm$ 200 & 5.29 $\pm$ 0.08 & 1.0 $\pm$ 0.1 & 9000 $\pm$ 300 & 8700 $\pm$ 300 & Component Aa\\
Ab & 2 & 5.488 $\pm$ 0.002 & 9700 $\pm$ 700 & 5.2 $\pm$ 0.2 & 1.0 $\pm$ 0.3 & 9900 $\pm$ 900 & 9000 $\pm$ 1000 & Component Aa\\
Ab & 3 & 10.977 $\pm$ 0.004 & 9155 $\pm$ 600 & 5.0 $\pm$ 0.3 & 0.4 $\pm$ 0.3 & 9600 $\pm$ 900 & 8700 $\pm$ 900 & Component Aa\\
Ab & 4 & 5.482 $\pm$ 0.002 & 7700 $\pm$ 700 & 4.9 $\pm$ 0.3 & 0.4 $\pm$ 0.3 & 8000 $\pm$ 1000 & 8000 $\pm$ 1000 & Component Aa\\
Ab & 5 & 5.488 $\pm$ 0.002 & 9500 $\pm$ 600 & 5.5 $\pm$ 0.2 & 1.4 $\pm$ 0.3 & 8200 $\pm$ 700 & 10100 $\pm$ 800 & Component Aa\\
Ab & 6 & 5.489 $\pm$ 0.002 & 9600 $\pm$ 700 & 4.9 $\pm$ 0.2 & 0.6 $\pm$ 0.2 & 10000 $\pm$ 1000 & 9000 $\pm$ 1000 & Component Aa\\
Ab & 8 & 5.492 $\pm$ 0.002 & 7600 $\pm$ 900 & 4.2 $\pm$ 0.3 & 0.4 $\pm$ 0.3 & 8000 $\pm$ 1000 & 8000 $\pm$ 800 & Component Aa\\
Ab & 9 & 5.489 $\pm$ 0.001 & 8900 $\pm$ 600 & 5.1 $\pm$ 0.2 & 0.4 $\pm$ 0.2 & 9300 $\pm$ 700 & 8500 $\pm$ 800 & Component Aa\\
Ab & 10 & 5.488 $/pm$ 0.002 & 9002 $\pm$ 700 & 5.1 $\pm$ 0.3 & 0.8 $\pm$ 0.3 & 9500 $\pm$ 700 & 7000 $\pm$ 1000 & Component Aa\\
Ab & 11 & 5.485 $\pm$ 0.001 & 8000 $\pm$ 500 & 5.0 $\pm$ 0.3 & 0.6 $\pm$ 0.3 & 8400 $\pm$ 800 & 800 $\pm$ 800 & Component Aa\\
Ab & 12 & 5.485 $\pm$ 0.002 & 9600 $\pm$ 800 & 5.0 $\pm$ 0.3 & 0.8 $\pm$ 0.3 & 10000 $\pm$ 1000 & 10000 $\pm$ 1000 & Component Aa\\
\midrule
B & 1-6 & 2.83721 $\pm$ 0.00002 & 26000 $\pm$ 300 & 3.34 $\pm$ 0.04 & 1.67 $\pm$ 0.02 & 16700 $\pm$ 400 & 33900 $\pm$ 400 & n/a\\
B & 1-9 & 2.83718 $\pm$ 0.00001 & 25400 $\pm$ 300 & 3.34 $\pm$ 0.04 & 1.67 $\pm$ 0.02 & 16700 $\pm$ 300 & 33500 $\pm$ 300 & n/a\\
B & 1-13 & 2.837163 $\pm$ 0.000007 & 25100 $\pm$ 200 & 3.34 $\pm$ 0.03 & 1.67 $\pm$ 0.02 & 16400 $\pm$ 300 & 33400 $\pm$ 300 & n/a\\
B & 1-13* & 2.837162 $\pm$ 0.000007 & 25100 $\pm$ 200 & 3.34 $\pm$ 0.03 & 1.67 $\pm$ 0.02 & 16400 $\pm$ 200 & 33400 $\pm$ 300 & n/a\\
B & 3 & 2.8372 $\pm$ 0.0004 & 28400 $\pm$ 700 & 3.3 $\pm$ 0.1 & 1.65 $\pm$ 0.05 & 34400 $\pm$ 900 & 17000 $\pm$ 1000 & n/a\\
B & 4 & 2.8366 $\pm$ 0.0004 & 23700 $\pm$ 800 & 3.3 $\pm$ 0.1 & 1.63 $\pm$ 0.06 & 18600 $\pm$ 900 & 33000 $\pm$ 1000 & n/a\\
B & 5 & 2.8373 $\pm$ 0.0003 & 26500 $\pm$ 600 & 3.31 $\pm$ 0.08 & 1.65 $\pm$ 0.04 & 33800 $\pm$ 700 & 15800 $\pm$ 800 & n/a\\
B & 8 & 2.8370 $\pm$ 0.0003 & 23000 $\pm$ 1000 & 3.4 $\pm$ 0.1 & 1.71 $\pm$ 0.06 & 31000 $\pm$ 1000 & 17000 $\pm$ 1000 & n/a\\
B & 9 & 2.8369 $\pm$ 0.0003 & 23400 $\pm$ 700 & 3.3 $\pm$ 0.1 & 1.66 $\pm$ 0.05 & 16600 $\pm$ 900 & 32600 $\pm$ 900 & n/a\\
B & 11 & 2.8371 $\pm$ 0.0003 & 23800 $\pm$ 700 & 3.3 $\pm$ 0.1 & 1.63 $\pm$ 0.05 & 32300 $\pm$ 800 & 16400 $\pm$ 900 & n/a\\
\midrule
Ba & 1 & 5.6742 $\pm$ 0.0009 & 37000 $\pm$ 1000 & 3.4 $\pm$ 0.1 & 1.70 $\pm$ 0.05 & 39000 $\pm$ 2000 & 34000 $\pm$ 2000 & 15000 $\pm$ 1000\\
Ba & 1-2 & 5.6744 $\pm$ 0.0002 & 35000 $\pm$ 700 & 3.32 $\pm$ 0.07 & 1.66 $\pm$ 0.04 & 36800 $\pm$ 900 & 33000 $\pm$ 2000 & Component Bb\\
Ba & 1-3 & 5.6744 $\pm$ 0.0001 & 34700 $\pm$ 500 & 3.29 $\pm$ 0.06 & 1.65 $\pm$ 0.03 & 35600 $\pm$ 700 & 33600 $\pm$ 800 & Component Bb\\
Ba & 2 & 5.6742 $\pm$ 0.0007 & 33000 $\pm$ 1000 & 3.21 $\pm$ 0.09 & 1.60 $\pm$ 0.04 & 33000 $\pm$ 1000 & 34000 $\pm$ 1000 & Component Bb\\
Ba & 6 & 5.6745 $\pm$ 0.0007 & 32600 $\pm$ 900 & 3.38 $\pm$ 0.09 & 1.69 $\pm$ 0.05 & 33000 $\pm$ 2000 & 33000 $\pm$ 1000 & Component Bb\\
Ba & 7 & 5.6736 $\pm$ 0.0007 & 33000 $\pm$ 1000 & 3.3 $\pm$ 0.1 & 1.64 $\pm$ 0.05 & 33000 $\pm$ 1000 & 34000 $\pm$ 1000 & Component Bb\\
Ba & 12 & 5.6740 $\pm$ 0.0008 & 36000 $\pm$ 1000 & 3.3 $\pm$ 0.1 & 1.67 $\pm$ 0.05 & 35000 $\pm$ 2000 & 36000 $\pm$ 2000 & Component Bb\\
Ba & 13 & 5.671 $\pm$ 0.0008 & 34000 $\pm$ 1000 & 3.3 $\pm$ 0.1 & 1.65 $\pm$ 0.06 & 36000 $\pm$ 2000 & 33000 $\pm$ 2000 & 2300 $\pm$ 900\\
\midrule
Bb & 1-2 & 5.6742 $\pm$ 0.0006 & 15800 $\pm$ 600 & 3.3 $\pm$ 0.2 & 1.64 $\pm$ 0.08 & 15000 $\pm$ 1000 & 16600 $\pm$ 900 & Component Ba\\
Bb & 1-3 & 5.6747 $\pm$ 0.0003 & 16000 $\pm$ 600 & 3.4 $\pm$ 0.2 & 1.67 $\pm$ 0.07 & 15000 $\pm$ 1000 & 16600 $\pm$ 700 & Component Ba\\
Bb & 2 & 5.675 $\pm$ 0.001 & 14900 $\pm$ 900 & 3.3 $\pm$ 0.2 & 1.64 $\pm$ 0.1 & 16000 $\pm$ 2000 & 14000 $\pm$ 1000 & Component Ba\\
Bb & 6 & 5.675 $\pm$ 0.001 & 16500 $\pm$ 700 & 3.3 $\pm$ 0.2 & 1.66 $\pm$ 0.09 & 18000 $\pm$ 1000 & 15000 $\pm$ 1000 & Component Ba\\
Bb & 7 & 2.836 $\pm$ 0.003 & 15000 $\pm$ 2000 & 3.3 $\pm$ 0.3 & 1.6 $\pm$ 0.2 & Model fitter failed & Model fitter failed & n/a\\
Bb & 10 & 5.6754 $\pm$ 0.0009 & 15400 $\pm$ 800 & 3.2 $\pm$ 0.2 & 1.6 $\pm$ 0.1 & 17000 $\pm$ 1000 & 14000 $\pm$ 1000 & 2331.4 $\pm$ 600\\
Bb & 12 & 5.6724 $\pm$ 0.0001 & 17000 $\pm$ 1000 & 3.1 $\pm$ 0.2 & 1.5 $\pm$ 0.1 & 15000 $\pm$ 2000 & 18000 $\pm$ 2000 & Component Ba\\
\bottomrule
\end{tabular}}
\end{table*}

\appendix{ETV tables}
\label{Appendix:ETVtab}

\begin{table}
\caption{Times of minima of TIC 278956474A. The first four items give seasonal minima deduced from WASP-South observations. Other data refer to individual eclipses observed by \textit{TESS} spacecraft. Integer and half-integer cycle numbers refer to primary and secondary eclipses. For the ETV analysis, discussed in the main part of the paper, only primary eclipses were used. Five primary eclipses marked with asterisks were omitted from the analysis as outliers. }
 \label{Tab:TIC_278956474A_ToM}
\centering
\begin{tabular}{@{}lrllrllrl}
\toprule
BJD & Cycle  & std. dev. & BJD & Cycle  & std. dev. & BJD & Cycle  & std. dev. 
\\ 
$-2\,400\,000$ & no. &   \multicolumn{1}{c}{$(d)$} & $-2\,400\,000$ & no. &   
\multicolumn{1}{c}{$(d)$} & $-2\,400\,000$ & no. &   \multicolumn{1}{c}{$(d)$} 
\\ 
\midrule
54826.636808 & -638.0 & 0.000549 & 58445.944896 &   21.5 & 0.001468 & 
58572.167299 &   44.5 & 0.000979 \\
55194.333657 & -571.0 & 0.001098 & 58448.697063 &   22.0 & 0.000144 & 
58574.921743 &   45.0 & 0.000141 \\
55562.031003 & -504.0 & 0.001098 & 58454.185939*&   23.0 & 0.000128 & 
58577.656449 &   45.5 & 0.000872 \\ 
55924.227778 & -438.0 & 0.001098 & 58456.919263 &   23.5 & 0.001334 & 
58580.409879 &   46.0 & 0.000133 \\ 
58327.962429 &    0.0 & 0.000174 & 58459.673021 &   24.0 & 0.000130 & 
58585.899280*&   47.0 & 0.000168 \\ 
58330.700768 &    0.5 & 0.002383 & 58462.411516 &   24.5 & 0.002015 & 
58588.633820 &   47.5 & 0.001724 \\ 
58333.449810 &    1.0 & 0.000223 & 58470.649037 &   26.0 & 0.000121 & 
58591.386912 &   48.0 & 0.000130 \\ 
58336.192167 &    1.5 & 0.001068 & 58473.389849 &   26.5 & 0.001229 & 
58594.119542 &   48.5 & 0.000951 \\ 
58341.672992 &    2.5 & 0.002330 & 58476.136447 &   27.0 & 0.000126 & 
58599.627581 &   49.5 & 0.004254 \\ 
58344.425319 &    3.0 & 0.000195 & 58478.879249 &   27.5 & 0.001043 & 
58602.362426 &   50.0 & 0.000175 \\ 
58347.162986 &    3.5 & 0.001927 & 58481.624776 &   28.0 & 0.000129 & 
58605.100888 &   50.5 & 0.001569 \\ 
58349.913744 &    4.0 & 0.000150 & 58484.363435 &   28.5 & 0.000751 & 
58607.850858 &   51.0 & 0.000153 \\ 
58352.658427 &    4.5 & 0.003649 & 58487.112762 &   29.0 & 0.000132 & 
58613.338684 &   52.0 & 0.000178 \\ 
58355.401627 &    5.0 & 0.000131 & 58489.871693 &   29.5 & 0.001835 & 
58616.075255 &   52.5 & 0.001549 \\ 
58358.135368 &    5.5 & 0.000848 & 58492.600453 &   30.0 & 0.000156 & 
58618.827111 &   53.0 & 0.000147 \\ 
58360.890011 &    6.0 & 0.000140 & 58495.332830 &   30.5 & 0.001338 & 
58621.561695 &   53.5 & 0.002102 \\ 
58363.626999 &    6.5 & 0.000658 & 58498.091166*&   31.0 & 0.000114 & 
58627.044601 &   54.5 & 0.001205 \\ 
58366.377135 &    7.0 & 0.000161 & 58500.833321 &   31.5 & 0.001364 & 
58629.803406 &   55.0 & 0.000155 \\ 
58369.122007 &    7.5 & 0.001326 & 58506.312484 &   32.5 & 0.001090 & 
58632.542077 &   55.5 & 0.000921 \\ 
58371.864689 &    8.0 & 0.000139 & 58509.064706 &   33.0 & 0.000139 & 
58635.290572 &   56.0 & 0.000146 \\ 
58374.600726 &    8.5 & 0.001342 & 58511.819107 &   33.5 & 0.001813 & 
58638.019856 &   56.5 & 0.001641 \\ 
58377.352284 &    9.0 & 0.000176 & 58514.552784 &   34.0 & 0.000109 & 
58640.779257 &   57.0 & 0.000178 \\ 
58380.083629 &    9.5 & 0.001641 & 58520.041266 &   35.0 & 0.000131 & 
58643.516085 &   57.5 & 0.001110 \\ 
58388.328366 &   11.0 & 0.000168 & 58522.785816 &   35.5 & 0.001032 & 
58646.267602 &   58.0 & 0.000178 \\ 
58391.068826 &   11.5 & 0.001415 & 58525.529251 &   36.0 & 0.000152 & 
58648.989799 &   58.5 & 0.001197 \\ 
58393.816827 &   12.0 & 0.000204 & 58528.268281 &   36.5 & 0.001249 & 
58651.755432 &   59.0 & 0.000158 \\ 
58399.305162 &   13.0 & 0.000119 & 58536.507117*&   38.0 & 0.000140 & 
58654.489005 &   59.5 & 0.001140 \\ 
58402.042463 &   13.5 & 0.002526 & 58539.236620 &   38.5 & 0.002777 & 
58657.244598 &   60.0 & 0.000171 \\ 
58404.792220 &   14.0 & 0.000199 & 58541.983305*&   39.0 & 0.000730 & 
58659.983219 &   60.5 & 0.000950 \\ 
58413.012446 &   15.5 & 0.001254 & 58544.732586 &   39.5 & 0.001435 & 
58662.733098 &   61.0 & 0.000206 \\ 
58415.768382 &   16.0 & 0.000163 & 58547.480518 &   40.0 & 0.000138 & 
58665.481953 &   61.5 & 0.001844 \\ 
58426.744932 &   18.0 & 0.000142 & 58550.226769 &   40.5 & 0.002788 & 
58670.967933 &   62.5 & 0.001479 \\ 
58429.484915 &   18.5 & 0.004152 & 58552.968752 &   41.0 & 0.000133 & 
58673.708794 &   63.0 & 0.000138 \\ 
58432.234014*&   19.0 & 0.000134 & 58558.457413 &   42.0 & 0.000118 & 
58676.450537 &   63.5 & 0.001335 \\ 
58434.963124 &   19.5 & 0.002461 & 58561.201168 &   42.5 & 0.001212 & 
58679.197053 &   64.0 & 0.000169 \\ 
58440.456132 &   20.5 & 0.001355 & 58563.945780 &   43.0 & 0.000125 & 
58681.947227 &   64.5 & 0.001120 \\ 
58443.208790 &   21.0 & 0.000129 & 58566.683307 &   43.5 & 0.001859 &  &    &  
\\ 
\bottomrule
\end{tabular}
\end{table}

\begin{table}
\caption{Times of minima of TIC 278956474B. Integer and half-integer cycle numbers refer to primary and secondary eclipses. For the ETV analysis, discussed in the main part of the paper, only primary eclipses were used. Primary eclipses marked with asteriskswere omitted from the analysis as outliers.}
 \label{Tab:TIC_278956474B_ToM}
\centering
\begin{tabular}{@{}lrllrllrl}
\toprule
BJD & Cycle  & std. dev. & BJD & Cycle  & std. dev. & BJD & Cycle  & std. dev. 
\\ 
$-2\,400\,000$ & no. &   \multicolumn{1}{c}{$(d)$} & $-2\,400\,000$ & no. &   
\multicolumn{1}{c}{$(d)$} & $-2\,400\,000$ & no. &   \multicolumn{1}{c}{$(d)$} 
\\ 
\midrule
58327.851141 &   -0.5 & 0.001078 & 58452.679392 &   21.5 & 0.001835 & 
58571.848064 &   42.5 & 0.004826 \\ 
58330.686768 &    0.0 & 0.000472 & 58455.526205 &   22.0 & 0.000464 & 
58574.684476 &   43.0 & 0.000540 \\ 
58333.528644 &    0.5 & 0.024207 & 58458.360384 &   22.5 & 0.001879 & 
58577.518764 &   43.5 & 0.000973 \\ 
58336.361373 &    1.0 & 0.000515 & 58461.196935*&   23.0 & 0.000365 & 
58580.356649 &   44.0 & 0.000601 \\ 
58342.033623*&    2.0 & 0.000564 & 58464.050752 &   23.5 & 0.084498 & 
58586.031380 &   45.0 & 0.000574 \\ 
58344.876998 &    2.5 & 0.001018 & 58469.709234 &   24.5 & 0.000666 & 
58588.873801 &   45.5 & 0.001518 \\ 
58347.706677*&    3.0 & 0.001567 & 58472.547953 &   25.0 & 0.000955 & 
58591.706607 &   46.0 & 0.000473 \\ 
58350.555515 &    3.5 & 0.001240 & 58475.388524 &   25.5 & 0.000929 & 
58594.549605 &   46.5 & 0.000516 \\ 
58356.227407 &    4.5 & 0.000779 & 58478.217404*&   26.0 & 0.000962 & 
58600.216707 &   47.5 & 0.001307 \\ 
58359.060381 &    5.0 & 0.000372 & 58481.055791 &   26.5 & 0.001051 & 
58603.056249 &   48.0 & 0.000425 \\ 
58361.901701 &    5.5 & 0.000839 & 58483.891625*&   27.0 & 0.000601 & 
58605.887612 &   48.5 & 0.000994 \\ 
58364.732752 &    6.0 & 0.000722 & 58486.738529 &   27.5 & 0.001463 & 
58608.722728*&   49.0 & 0.000537 \\ 
58370.408184 &    7.0 & 0.000478 & 58489.570680 &   28.0 & 0.000380 & 
58614.401286 &   50.0 & 0.000479 \\ 
58373.248194 &    7.5 & 0.001198 & 58492.406297 &   28.5 & 0.001592 & 
58617.242303 &   50.5 & 0.001257 \\ 
58376.082783 &    8.0 & 0.000501 & 58495.246523 &   29.0 & 0.000561 & 
58620.073963*&   51.0 & 0.000724 \\ 
58378.922654 &    8.5 & 0.001383 & 58498.085931 &   29.5 & 0.000965 & 
58622.913595 &   51.5 & 0.001044 \\ 
58387.429007*&   10.0 & 0.000441 & 58500.914410*&   30.0 & 0.000567 & 
58625.754045 &   52.0 & 0.000944 \\ 
58390.269093 &   10.5 & 0.003915 & 58506.594229 &   31.0 & 0.000351 & 
58628.598291 &   52.5 & 0.001760 \\ 
58393.105081 &   11.0 & 0.000456 & 58509.437963 &   31.5 & 0.001246 & 
58631.428199 &   53.0 & 0.000505 \\ 
58398.781892 &   12.0 & 0.000383 & 58512.267718 &   32.0 & 0.000517 & 
58634.269473 &   53.5 & 0.001228 \\ 
58401.621516 &   12.5 & 0.001218 & 58515.106161 &   32.5 & 0.000786 & 
58637.101893 &   54.0 & 0.000492 \\ 
58404.454889 &   13.0 & 0.000390 & 58517.935775*&   33.0 & 0.000618 & 
58642.776264 &   55.0 & 0.000615 \\ 
58412.965130 &   14.5 & 0.001871 & 58520.781624 &   33.5 & 0.001152 & 
58645.618682 &   55.5 & 0.000989 \\ 
58415.799460*&   15.0 & 0.000477 & 58523.616511 &   34.0 & 0.000617 & 
58648.446307 &   56.0 & 0.000439 \\ 
58421.478895 &   16.0 & 0.002228 & 58526.449138 &   34.5 & 0.000690 & 
58651.283550 &   56.5 & 0.001568 \\ 
58427.152712 &   17.0 & 0.000595 & 58537.802322 &   36.5 & 0.001126 & 
58654.122005 &   57.0 & 0.000305 \\ 
58429.994122 &   17.5 & 0.001210 & 58540.637289 &   37.0 & 0.000328 & 
58656.962553 &   57.5 & 0.001599 \\ 
58432.827096 &   18.0 & 0.000552 & 58546.311254 &   38.0 & 0.000428 & 
58659.797199 &   58.0 & 0.001004 \\ 
58435.667600 &   18.5 & 0.000915 & 58549.151810 &   38.5 & 0.000682 & 
58662.627503 &   58.5 & 0.001251 \\ 
58438.498776*&   19.0 & 0.000330 & 58551.986126 &   39.0 & 0.000410 & 
58665.470728 &   59.0 & 0.000598 \\ 
58441.334196 &   19.5 & 0.001504 & 58554.816386 &   39.5 & 0.001820 & 
58671.145005 &   60.0 & 0.000736 \\ 
58444.173601*&   20.0 & 0.000600 & 58560.498489 &   40.5 & 0.000809 & 
58673.988971 &   60.5 & 0.001242 \\ 
58447.015592 &   20.5 & 0.001109 & 58563.336764 &   41.0 & 0.000548 & 
58676.819757 &   61.0 & 0.000798 \\ 
58449.850482 &   21.0 & 0.000593 & 58566.164903 &   41.5 & 0.001345 & 
58679.659708 &   61.5 & 0.002367 \\ 
\bottomrule
\end{tabular}
\end{table}

\end{document}